\documentclass{aa}
\usepackage{graphicx}
\usepackage{txfonts}

\begin{document}
   \title{Starbursts in isolated galaxies.
          I. The influence of stellar birth function and IMF}

   \subtitle{}

   \author{Ch.\ Theis
           \inst{1,4}
          \and
           J. K\"oppen\inst{2,3,4}
          }

%   \offprints{Ch.\ Theis}

   \institute{Institute of Astronomy, University of Vienna,
              T\"urkenschanzstr.\ 17, A-1180 Vienna, Austria\\
              \email{theis@astro.univie.ac.at}
         \and    
              Observatoire Astronomique de Strasbourg, 11, rue de
                  l'Universit\'e, F--67000 Strasbourg, France\\
              \email{koppen@astro.u-strasbg.fr}
         \and    
              International Space University, Parc d'Innovation,
              1, rue Jean-Dominique Cassini,
              F--67400 Illkirch-Graffenstaden, France
         \and    
              Institut f\"ur Theoretische Physik und Astrophysik, 
              Universit\"at Kiel, D--24098 Kiel, Germany
             }

   \date{Received / Accepted }

   \abstract
%
%   context
%
   {Starbursts and substantial variations in the star formation
     histories are a common phenomenon in galaxies. Although
     predominantly found in interacting galaxies, they also occur in
     isolated galaxies.}
%
%   aims
%
   {We study the stability properties of isolated star-forming dwarf
     galaxies with the aim of identifying starburst modes.  The impact of
     the stellar birth function, i.e.\ a spontaneous and an
     induced star formation mode, the initial mass function (IMF), the
     stellar feedback and the interstellar medium (ISM) model on the
     galactic star formation history are investigated. We especially
     focus on dynamically driven starbursts induced by stellar
     feedback.}
%
%   methods
%
{We apply a one-zone model for a star-gas system coupled by both mass
  and energy transfer. Additionally, we extend the network for
  active dynamical evolution.  This allows for a coupling between the
  dynamical state of the galaxy and its internal properties, such as star
  formation activity or the thermal state of the ISM.}
%
%  results
%
{While the influence of the dynamics on the total star formation rate
  is strong, especially with nonlinear stellar birth functions, the
  coupling of the internal properties (gas temperature) on the
  dynamics is rather limited, because radiative cooling keeps the gas
  temperature well below the virial temperature. Because of short cooling
  and feedback timescales, the star formation rate is close to the
  equilibrium star formation rates.  Quasi-periodic starbursts occur,
  because star formation follows the variations in the gas density
  induced by decaying virial oscillations. This behaviour is quite
  insensitive to the nature and the details of the stellar birth
  description, viz.\ whether spontaneous or induced star formation 
  is considered or the IMF is varied. A second type of burst is found as
  an instability operating when the cooling may drop at very low densities
  with increasing temperature, in regimes beyond $10^4$ K.  }
%
% conclusions
%
{Bursts of star formation occur during transitory phases, when
  dynamical equilibrium is established. Then they are quasi-periodic
  on the dynamical timescale. Because of short heating and cooling
  timescales, the star formation rate follows the equilibrium star formation rate
  corresponding to the actual gas density.}

   \maketitle

%##############################################
%              Introduction
%##############################################

\section{Introduction}
\label{sect_introduction}

   Variations in the star formation rate (SFR) in dwarf galaxies have
been deduced from several observations like direct measurements of
individual star formation histories (Dohm-Palmer et al.\ 
\cite{dohmpalmer02}), a large variation in the specific SFR, or short
gas depletion times (Gallagher \& Hunter \cite{gallagher84}).
Especially interesting is the group of isolated starburst galaxies,
because no external triggers can be invoked for their high activity.
Van Zee (\cite{vanzee01}) studied a large sample of isolated dwarf
irregular galaxies. For the majority of these galaxies, she found a
fairly constant SFR over the past 10 Gyrs indicating
that they reached a quasi-equilibrium state burning on a long gas
depletion time of about 20 Gyr. However, a small fraction of her
sample shows high star formation activity that is strongly
concentrated in the central region and/or characterised by small
radial scale lengths, i.e.\ enhanced gas densities. Already Gallagher
\& Hunter (\cite{gallagher84}) stressed that the observation of
starburst dIrrs implies a mechanism to organise large-scale 
star formation which cannot operate all the time due to fast
consumption of fuel. On the other hand, the instability leading to an
episodic SFR should only work in a small parameter range, because most
of the isolated dIrrs seem to evolve in a highly self-regulated
manner.

The first theoretical models dealing with star formation variations in
galaxies were based on closed-box models. For reasonable models of the
interstellar medium (ISM) it turned out that stellar feedback is very
effective in suppressing instabilities of the ISM (e.g.\ Ikeuchi \&
Tomita \cite{ikeuchi83}, Scalo \& Struck-Marcell \cite{scalo86}).
These models only allow for a burst-like behaviour when a
long time-delay (of
  the order of $10^8$ yrs) between star formation and stellar feedback is
introduced. Recently Quillen \& Bland-Hawthorn (\cite{quillen08})
emphasised the importance of such a long time-delay of the feedback
for creating episodic star formation in Milky Way type galaxies.
The drawback of these models is the long time-delay that
  exceeds the short feedback timescales related to massive star
  formation by far. The latter are a few Myr at most
  (if not quasi-instantaneously when taking stellar winds into
  account). It is not clear what has kept the ISM from experiencing
  feedback by stars for such a long time.

K\"oppen et al.\ (\cite{koeppen95}, hereafter KTH95) investigated a
simple numerical model motivated by full chemodynamical simulations.
They show that the effective SFR is almost
independent of the detailed recipe for the stellar birth function, 
provided a negative feedback due to the thermal state of the ISM
is considered in the stellar birth function. This also holds for a multi-phase ISM
(K\"oppen et al.\ \cite{koeppen98}). All these box models suffer,
however, from neglecting galactic dynamics.  Additionally, they are
designed for small galactic regions of a few 100 pc.
As a result, coherence between different unstable regions and global
starbursts could not be modelled.
%Hence, coherence between different unstable regions is not obvious, by
%this missing a {\it global} starburst.

Recently, several attempts have been made to study the evolution of
isolated dwarf galaxies by taking full stellar and gas dynamics, as
well as star formation and stellar feedback, into account (e.g.\
Hensler et al.\ \cite{hensler04}, Pelupessy et al.\
\cite{pelupessy04}, \cite{struck05}).  Though these calculations
differ in many details, they agree in producing large-scale star formation
variations on dynamical timescales in some of their models.  This
indicates the crucial role dynamics plays for episodic starburst
behaviour.

Starbursts are often associated with galaxy interactions. E.g.\ most
of the ultraluminous infrared galaxies are interpreted as the result
of galaxy interactions or mergers (e.g.\ Sanders \& Mirabel
\cite{sanders96}). It is interesting that the response of galaxies to
interaction-induced perturbations is not unique or simple. Di Matteo
et al.\ (\cite{dimatteo07}) present a subset of merger events
created in a numerical survey of galaxy collisions.  Their Figs.\ 8
and 9 show a variety of reaction patterns of the SFR after the first
closest approach, ranging from strong starbursts to almost no
enhancement in the star formation activity. 
%The simple statement ''galaxy
%interaction leads to starbursts'' does not necessarily hold.
Another example for dwarf galaxy interactions is presented in
Bekki (\cite{bekki08}).
Despite this model-to-model variation, statistics of large 
  galaxy-pair samples show a clear correlation between interactions and star
  formation activity. For example using 12500 galaxy pairs from the SDSS
  survey Nikolic et al.\ (\cite{nikolic04}) find an enhanced star
  formation rate for projected separations below 50 kpc and an
  anti-correlation between the projected separation and the star
  formation rate (for more details, cf.\ to the review
  by Struck \cite{struck06}).

A major disadvantage of the simulations -- either for isolated or
interacting galaxies -- is their complexity.  This prevents a detailed
investigation of the parameter space and sometimes also hides the key
physical mechanisms. Instead of analysing dwarf galaxies by such
detailed and complex models, we study here a set of equations that is as simple as possible
that incorporates the main generic features of the
complex numerical simulations. The advantage is a (semi-)analytical
treatment and a more direct physical interpretation of the implemented
physics. Moreover, extended parameter studies can be carried out much more easily
with our set of equations than with the full numerical
modelling.

This paper introduces an extended version of
the analytical models of KTH95 that allows us to deal with the dynamical
evolution of a galaxy. By this, the evolution of the star formation
activity not only depends on the internal properties of the ISM,
but also on the dynamical state of the galaxy. As an application of our analytical
model, the star formation activity for dwarf galaxies in the mass regime of a few
$10^9 M_\odot$ is studied.

In this paper we investigate the influence on the star formation 
by studying a generalized stellar birth function (composed of a
spontaneous star formation mode with negative thermal feedback and 
an induced star formation mode), the initial stellar mass function,
and the stellar feedback.

This paper is organised as follows. In Sect.\ \ref{sect_numericalmodel} 
we introduce our set of equations. A reference model and the result
of parameter variations are presented in Sect.\ \ref{sect_results}. 
Section \ref{sect_discussion} contains a discussion of the results.

%##############################################
%            The numerical model
%##############################################

\section{The numerical model}
\label{sect_numericalmodel}

\subsection{The non-dynamical closed box model}

  Motivated by the complex and computationally expensive chemo-dynamical
calculations of Theis et al.\ (\cite{theis92}) (also used in
the simulations for dwarf galaxies by Hensler et al.\ 
(\cite{hensler04})), KTH95 analysed a set
of closed-box models that was designed to be as simple as possible,
but still keeps the main qualitative features of the chemo-dynamical
network (e.g.\ feedback processes). These simplified equations
describe the temporal evolution of the density of short-lived massive
stars ($s$), of gas ($g$), and of the internal energy ($e$) of the gas:
\begin{eqnarray}
    {d g \over dt} & = & -\Psi_b(g,T)   + \frac{\eta s}{\tau}
          \label{dgdtsm1} \\
    {d s \over dt} & = & \xi\Psi_b(g,T) -  \frac{s}{\tau}
          \label{dsdtsm1} \\
    {d e \over dt} & = & h(g)s  - g^2\Lambda(T) \,.
          \label{dedtsm1}
%    {d r \over dt} & = & (1-\xi)\Psi_b(g,T)  + (1-\eta)s/\tau \punkt
\end{eqnarray}
The gas is consumed by star formation denoted by a {\it stellar birth
function} $\Psi_b(g,T)$ depending on gas density $g$ and temperature
$T$\footnote{Because the thermal state of all gas components is
summarised in a single temperature $T$, this temperature is 
a ''typical'' temperature of the gas characterising the thermal energy or 
pressure of the ISM. This notion is rather similar e.g.\
to the notion of the external pressure in Blitz \& Rosolowsky (\cite{blitz04}).
}. The temperature is related to the energy density by $e=b g T$
with $b\equiv 3/2 \cdot k / \mu m_p$ ($k$ is the Boltzmann constant,
$\mu$ the mean molecular weight, and $m_p$ the proton mass). 
 The
second term in (\ref{dgdtsm1}) is a source term for the ISM due to
stellar winds and supernova ejecta replenishing the ISM. Massive stars
are assumed to return a fraction $\eta\approx90\%$ of their mass after
a mean stellar lifetime $\tau \approx 10 \, {\rm Myr}$. Similarly,
Eq.\ (\ref{dsdtsm1}) describes the formation of massive stars, as well
as their death ($\xi\approx 12\%$ is the mass fraction of
massive stars derived from the initial mass function (IMF)).

The last equation deals with the energy budget of the ISM gas.  Its
internal energy is characterised by its temperature $T$.  Assuming a
turbulence-driven ISM it receives heating mainly by massive stars
(e.g.\ Ly-continuum radiation or type II supernovae). The conversion
factor $h$ can be constant, if all the energy is absorbed, or it may
depend on gas density, e.g.\ for radiative heating in an optically
thin gas.  The main dissipative process of the gas is radiative
cooling.  The energy loss scales with $g^2$, where the proportionality
factor $\Lambda(T)$ is the classical cooling function.

The set of equations is closed by a generalized Schmidt-like stellar
birth function
\begin{equation}
   \label{psism1}
   \Psi_b(g,T) = \Psi_{b,{\rm sp}}(g,T) \equiv C_n \, g^n \cdot f(T) \,.
\end{equation}
This function corresponds to spontaneous star formation.  (An induced
star formation mode will be discussed later in Sect.\
\ref{sect_res_posfeedback}.) Different to the classical Schmidt-law we
allow for a (negative) thermal feedback
\begin{equation}
  f(T) \equiv e^{-T/T_s} \, ,
\end{equation}
which reduces the star formation efficiency in case of high gas
temperatures.\footnote{A negative temperature $T_s$ corresponds to a
  positive feedback. However, it is related to the gas temperature
  (velocity dispersion) alone. Therefore, we used
  a physically better motivated ansatz for positive stellar feedback
  in Sect.\ \ref{sect_res_posfeedback}.}
The exponent $n$ is varied between a linear ($n=1$) and a quadratic
($n=2$) Schmidt law; the constant $C_n$ is chosen to match the star
formation rate observed in the solar neighborhood. Setting $T_s$ to
infinity we recover the classical Schmidt law depending only on the
gas density.

In case of radiative cooling, thermal equilibrium is quickly
established due to short heating and cooling timescales (KTH95).  As a
result the effective SFR $\Psi_e$ (which allows for a
stationary solution of Eqs.\ (\ref{dsdtsm1}) and (\ref{dedtsm1})) is
independent of the adopted stellar birth function
\begin{equation}
   \label{eq_psieq}
   \Psi_e(g,T_e) = g^2 \frac{\Lambda(T_e)}{h(g) \xi \tau} \, ,
\end{equation}
where $T_e$ is the equilibrium gas temperature. First, it is remarkable
that $\Psi_e$ is between a linear and a quadratic Schmidt-law
depending on $h(g)$, but {\it not} depending on $\Psi_b$, especially
on the exponent $n$. Secondly, star formation depends almost
exclusively on gas density. This means that such systems undergo
starbursts only when the gas density is strongly enhanced, e.g.\ by
the result of a collapse, an infall, or an induced structural change
of the galaxy.
%As long as the cooling function increases with temperature, there is
%no intrinsic thermal instability leading to a burst. This is caused by
%the strong and quick stellar feedback.

\subsection{Including the dynamics}

   We extend our closed-box model for dynamical evolution by
considering a volume with characteristic radius $R$ 
(e.g.\ the half-mass radius) of the baryonic system of the galaxy. 
The equation of motion -- motivated by the corresponding equation for a thin 
gaseous shell -- is given by
\begin{eqnarray}
   \frac{d^2 R}{dt^2} & = & 
      \label{d2rdt2smd}
      - \frac{d\Phi_{\rm DM}}{dr}\Bigg|_{r=R} - \frac{1}{2} \frac{GM_b}{R^2}
      - \frac{1}{g} \cdot \frac{dP}{dr}
      + \frac{j^2}{R^3}
      - \frac{v_{\rm rad}}{\tau_{\rm fric}} \nonumber \\
   & = & 
     - \frac{d\Phi_{\rm DM}}{dr}\Bigg|_{r=R} - \frac{1}{2} \frac{GM_b}{R^2} \nonumber \\
   & &  + C_P \, \frac{k}{\mu m_p} \, \frac{T}{R} 
        + \frac{(C_j j_{\max})^2}{R^3}
        - \frac{v_{\rm rad}}{C_{\rm fric }\tau_{\rm ff}} \, .
\end{eqnarray}
The first two terms in (\ref{d2rdt2smd}) describe the gravity, i.e.\
the contribution of the dark matter potential and the self-gravity of
the baryons with the total mass $M_b$.  The third term estimates the
contribution of the pressure $P$.  The fourth term denotes the angular
momentum conservation where $j_{\rm max}$ is the maximum specific
angular momentum derived for a circular orbit at the initial radius
$R(t=0) = R_{\rm ini}$.  The last term describes decaying virial
oscillations as observed in collapse simulations, and $v_{\rm rad}$ is
the radial velocity $dR/dt$. The frictional timescale $\tau_{\rm fric}
= C_{\rm fric} \tau_{\rm ff}$ is normalised to the free-fall timescale
$\tau_{\rm ff}$ of the dark matter halo at the initial radius.

From the (half-mass) volume $V(t) \equiv 4\pi/3 \cdot R^3(t)$ of the
system and the mass of each component the densities can be derived,
e.g.\ the gas density reads
\begin{equation}
    g(t) = \frac{1}{2} \cdot \frac{M_{\rm g}(t)}{V(t)} \,.
\end{equation}
Integrating Eqs.\ (\ref{dgdtsm1}) - (\ref{dedtsm1}) over volume yields
the equations for the total masses of the gas, the massive stars, and
the total internal energy $E \equiv e V$ inside $R$:
\begin{eqnarray}
    {d M_g \over dt} & = & - \Psi(R,g,T)   + \frac{\eta M_s}{\tau}  
          \label{dgdtsmd} \\
    {d M_s \over dt} & = & \xi \, \Psi(R,g,T) -  \frac{M_s}{\tau}      
          \label{dsdtsmd} \\
%    {d E \over dt}   & = & h(g) M_s  - g^2 \Lambda(T) \cdot V(t)
%            - P \frac{dV}{dt}          
%          \label{dedtsmd} \\
      {d E \over dt}  & = & h(g) \, M_s  - \frac{E}{\tau_\mathrm{cool}(g,T)}
            - P \frac{dV}{dt} \, .
          \label{dedtsmd}
%    {d M_r \over dt} & = & (1-\xi)\Psi(R,g,T)  + (1-\eta) M_s / \tau \punkt
\end{eqnarray}

The total SFR $\Psi(R,g,T) \equiv \Psi_b(g,T) \cdot V$ (in
$\mathrm{M}_{\sun} \mathrm{yr}^{-1}$) depends not only on the ISM
parameters $g$ and $T$, but now also on the current size $R$ of the
system. It is worth noting that this applies only for a nonlinear
star formation law, because $\Psi(R,g,T) = \Psi_b(M_g/V,T) \cdot V
\propto M_g^n \cdot V^{1-n}$!

%{\bf XXX konsequenterweise m\"usste
%  hier eigentlich $\Psi \equiv 2 \Psi_b \cdot V$ stehen (wg.\ 
%  Halb-Massen-Radius); man kann den Faktor 2 allerdings in $C_n$
%  subsummieren oder das Ganze eh' nur als semi-quantitative Gleichung
%  ansehen wg.\ der vielen weiteren N\"aherungen. Im Programm ist
%  dieser Faktor 2 bislang unterschlagen \dots XXX}.

The temperature is related to the internal energy $E$ by $E = 1/2 \, b M_g T$.
The factor 1/2 stems from the assumption that $R$ is the half-mass radius. 
The last term in Eq.\ (\ref{dedtsmd}) denotes the $P dV$ work.
The pressure $P$ is given by $P = 2/3 \, E/V = 2/3 \, b g T$.  In case
of radiative dissipation, the corresponding timescale
$\tau_\mathrm{cool}(g,T)$ is defined as
\begin{equation}
   \tau_\mathrm{cool}(g,T) \equiv \frac{b T}{g \Lambda(T)} \, .
   \label{taurad}
\end{equation}
Dissipation by radiative cooling is included by a cooling
function combined from Dalgarno \& McCray (\cite{dalgarno72}) for
$T<10^4 {\rm K}$ and B\"ohringer \& Hensler (\cite{boehringer89}) for
higher temperatures (and a metallicity $Z=1/10\, Z_\odot$).  

For the dark matter halo, we apply the universal halo suggested by
Burkert (\cite{burkert95}) for dwarf galaxies. The scale radius $r_0$
of the Burkert halo was set to 8 kpc. For a baryonic mass of $M_{\rm
  gas} = 2 \cdot 10^{9} M_\odot$ and \mbox{$R_{\rm ini} = 8$ kpc} the
initial baryonic mass fraction is about 10\%.

\begin{figure}
  \includegraphics[height=.33\textheight,angle=270]{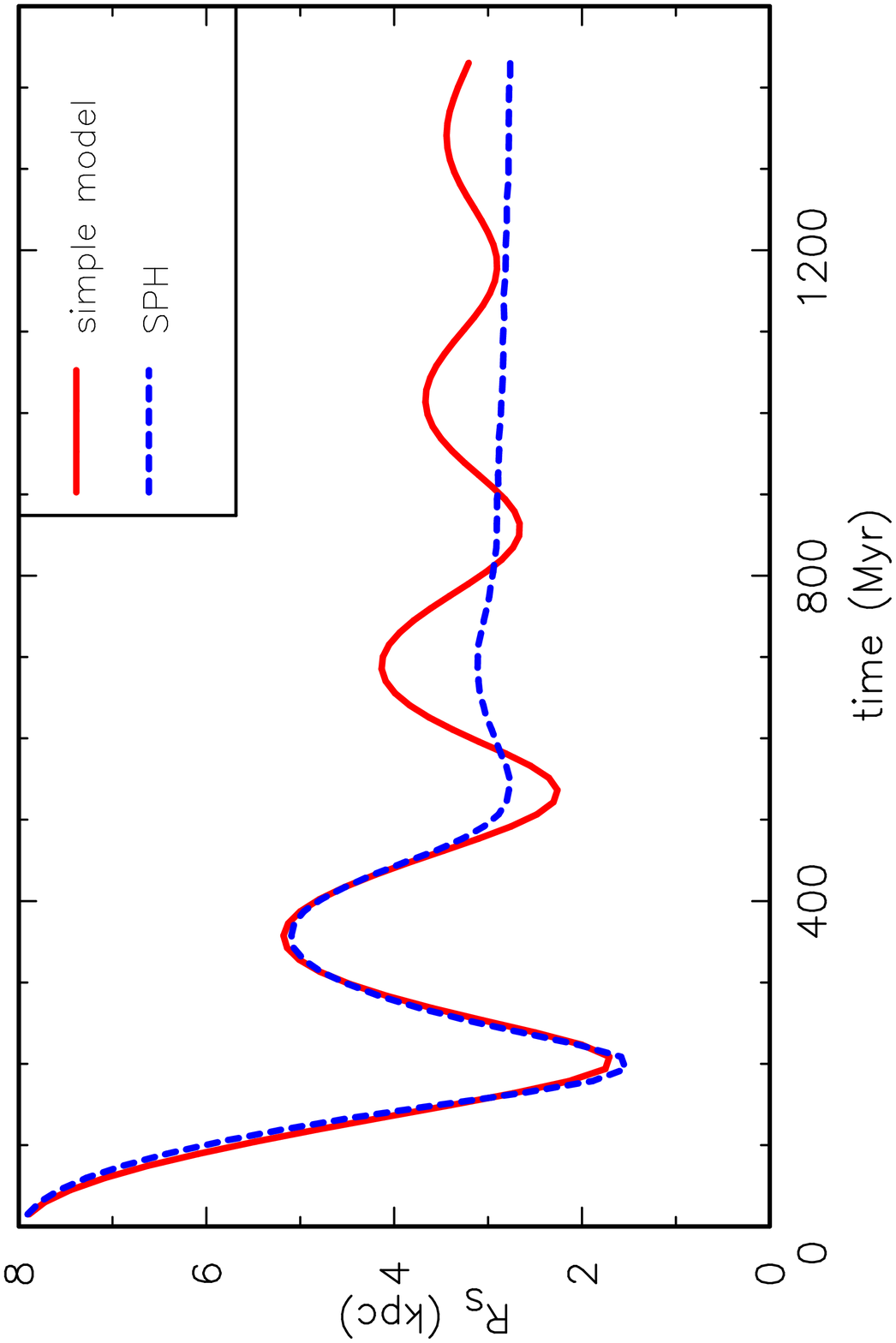}
  \includegraphics[height=.33\textheight,angle=270]{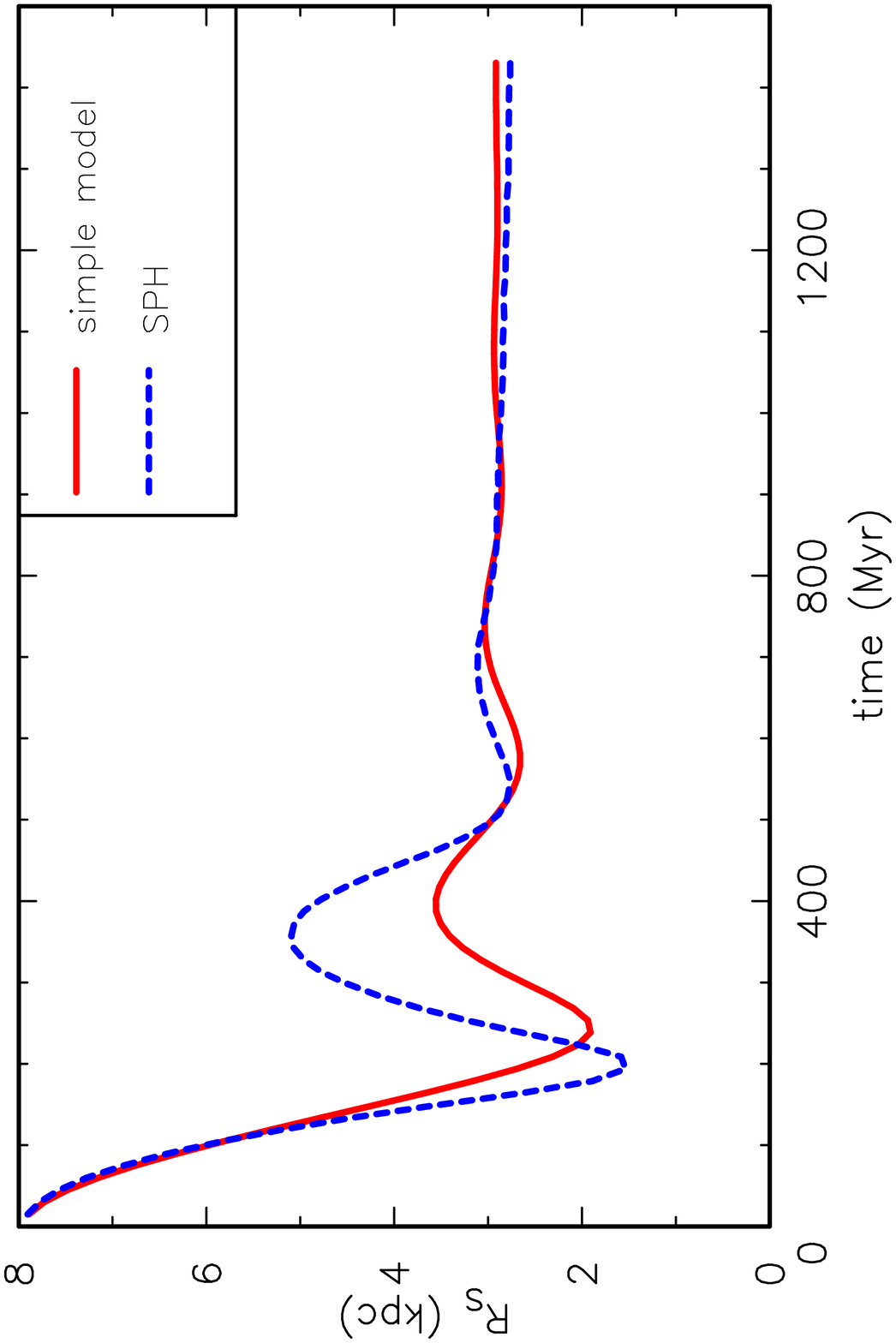}
  \caption{Temporal evolution of the half-mass radius in a 3D-SPH 
     simulation of a collapsing gas sphere in a dark matter halo and the result
     for the simple equation of motion (\ref{d2rdt2smd}). Shown are
     models fitting the early stage well (with $C_P=1.91, C_j=0.15, 
     C_f=2.06$, left) or the late stage (with $C_P=1.97, C_j=0.11, 
     C_f=0.88$, right).}
  \label{fig_cadjustment}
\end{figure}

The constants $C_P$, $C_j$, and $C_{\rm fric}$ are adjusted to a 3d-SPH
simulation of a collapsing gas sphere in a given static dark matter
halo potential. The simulations were done with the SPH part of
  the code described in Harfst et al.\ (\cite{harfst06}). Two sets of
constants were derived: set 1 ($C_P=1.91$, $C_j=0.15$, $C_{\rm
  fric}=2.06$) gives a good fit to the early evolution, whereas set 2
($C_P=1.97$, $C_j=0.11$, $C_{\rm fric}=0.88$) describes the late
evolution (Fig.\ \ref{fig_cadjustment}). The main freedom in the
choice of parameters comes from the damping behaviour that is covered
by a variation of a factor of 2 in the constant $C_{\rm fric}$, i.e.\
in the frictional timescale.  When minimising the deviations over all
periods the best fit was reached by $C_P=1.74$, $C_j=0.13$, $C_{\rm
  fric}=1.64$. If not stated differently, we apply these values.

The Eqs.\ (\ref{d2rdt2smd}) - (\ref{taurad}) yield a closed set of
equations including a first approximation for the dynamical evolution.
We solve the differential equations using a fourth-order Runge-Kutta
integrator with adaptive timestep (Press et al.\ \cite{press92}). The
integration accuracy is set to $10^{-10}$ (or better).  For
convenience we also consider an equation for the mass $M_r$ of the
long-lived low-mass stars and the stellar remnants
\begin{equation}
    {d M_r \over dt} = (1-\xi) \, \Psi(R,g,T) + (1-\eta) \frac{M_s}{\tau} \,.
    \label{drdtsmd}
\end{equation}
The formation of the low-mass stars is described by the first term and
the production of stellar remnants by the second term in Eq.\
(\ref{drdtsmd}).  We neglect the gas return by 
long-lived low-mass stars. This approximation is appropriate for gas-rich
systems. Therefore, our models cannot provide a realistic evolution of
the gas mass, if the gas fraction is smaller than a few percent.  From
Eqs.\ (\ref{dgdtsmd}) and (\ref{dsdtsmd}) we get the solution
\begin{equation}
  M_r(t) = M_\mathrm{b} - \left( M_g(t) + M_s(t) \right)
\end{equation}
describing mass conservation, and $M_\mathrm{b}$ is the total initial
mass in baryons.

A more accurate treatment would include mass return of low-mass stars
by planetary nebulae and type Ia supernovae. This would imply the
numerically rather expensive computation of the instantaneous mass
transfer rates from the details of the past star formation history for the
spectrum of stellar masses. For our purpose of studying the general
features of galaxy evolution with a simplified dynamics, such an
elaborate approach would not be such an advantage that it would justify the
computational effort.

%==============================
%           figure 1
%==============================

%---------------------------------------------------------
%            units
%---------------------------------------------------------

\subsection{Units}

    For our calculations we used the following units: 1 $M_{\sun}$, 1pc,
and 1 Myr. If not stated differently, parameters and results are given
in these units or the corresponding derived units, e.g.\ mass densities
in $M_{\sun} \, \mathrm{pc}^{-3}$.

%#############################################
%            Results
%#############################################

\section{Results}
\label{sect_results}

%  In this section we apply our set of equations to different ISM
%scenarios. In Sect.\ \ref{sect_res_diffuse_ism} we show the 
%evolution of a typical diffuse ISM model. Then we study the 
%influence of the parametrization of the stellar birth function
%(Sect.\ \ref{sect_res_sbf}).

%-------------------------------------------------
%       table of numerical models
%-------------------------------------------------

\begin{table}
  \caption{Properties of the numerical models}
  \label{table_models}
  \begin{tabular}{c|c|c}
    model & SF-law ($n$,$C_n$,$T_s$)  & comment \\ \hline
      A   & 1.5, 0.06,  $10^2$        & reference model  \\ \hline
      A2  & 1.5, 0.06,  $10^4$        & \\ \hline
    DN1   & 1.0, 0.007, $\infty$      & no feedback, Schmidt law \\
    DN2   & 1.1, 0.007, $\infty$      & \\
    DN3   & 1.5, 0.007, $\infty$      & \\
    DN4   & 2.0, 0.007, $\infty$      & \\
    DN5   & 2.5, 0.007, $\infty$      & \\ \hline
    DN6   & 2.0, 0.7,   $\infty$      & enhanced $C_n$ \\ \hline
    DT1   & 1.0, 0.007, $10^5$        & with thermal feedback \\
    DT2   & 1.0, 0.007, $10^4$        & \\
    DT3   & 1.0, 0.007, $10^3$        & \\
    DT4   & 1.0, 0.007, $10^2$        & \\ \hline
    DT5   & 2.0, 0.55,  $10^5$        & quadr.\ Schmidt law with feedback \\ \hline
    DI1   & ind.\ SF only             & induced star formation mode \\
    DI2   & model A, $T_s = 10^2$     & combined SF modes \\
    DI3   & model A, $T_s = 10^4$     & combined SF modes \\ \hline
    DX1   & model A, $\xi= 6\% $      & varied constant IMF \\
    DX2   & model A, $\xi= 12\%$      & \\ \hline
    DWK1  & 1.5, 0.06,  $10^2$        & Weidner-Kroupa IMF, SN heating \\
    DWK2  & 1.5, 0.06,  $10^2$        &   radiative heating \\ \hline
  \end{tabular}
\end{table}

%------------------------------------------------------------
%                The diffuse ISM models / reference model
%------------------------------------------------------------

\subsection{The reference model}
\label{sect_res_diffuse_ism}

%\subsection{The diffuse ISM models}
%\label{sect_res_diffuse_ism}

%==============================
%           figure 2
%==============================
%------------------------------
%       mod004b (early)
%------------------------------
\begin{figure*}
  \begin{center}
    \resizebox{0.9\hsize}{!}{
    \includegraphics[angle=270]{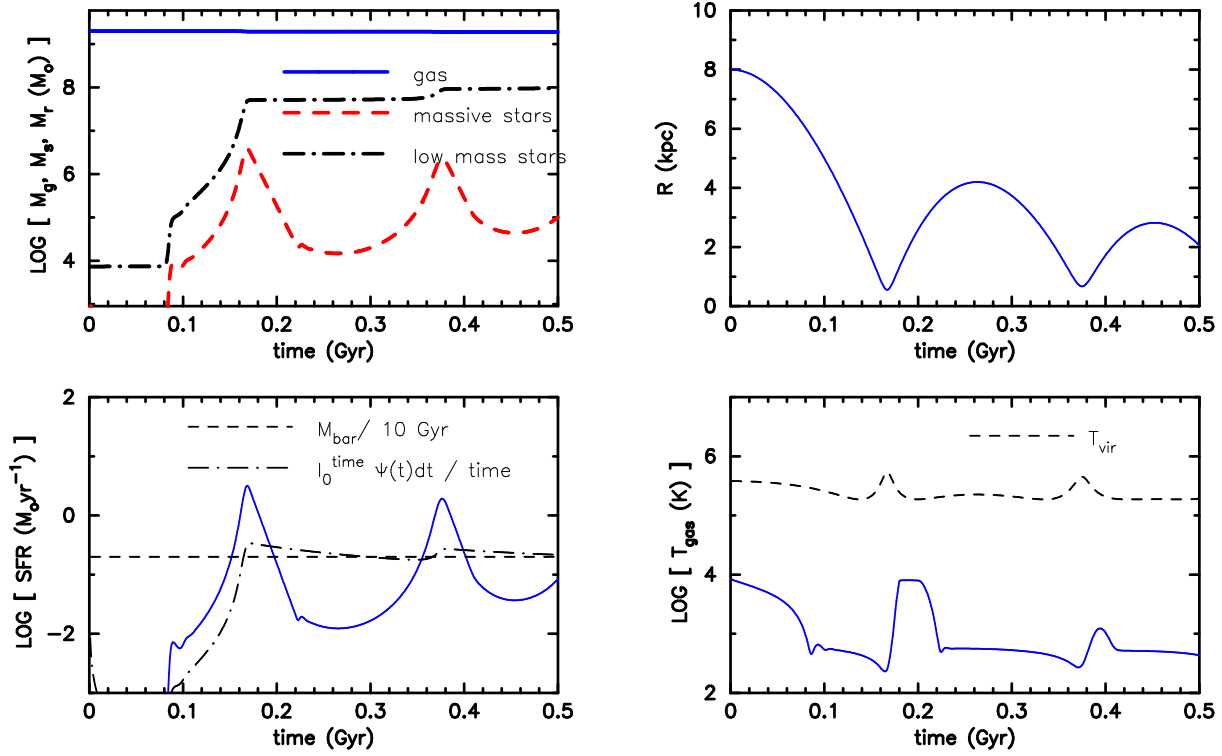}
     }
  \end{center}
  \caption{Initial temporal evolution of a model with a baryonic mass
    of $2 \cdot 10^9 \, {\rm M}_\odot$. Shown are the masses of the
    baryonic components (upper left), i.e.\ gas (solid), low-mass
    stars, and stellar remnants (dot-dashed), and massive stars
    (dashed), the mean radius of the system (upper right), the total
    SFR (lower left), and the gas temperatures (lower
    right).  The SFR is compared with the gas
    consumption averaged over 10 Gyrs and the mean star formation up to
    time $t$, i.e.\ $\int_0^t \Psi(t') dt' / t$.  The temperature is
    compared with the virial temperature of the system including the
    self-gravity of the baryons and the dark matter contribution.}
  \label{fig_ref}
\end{figure*}

%==============================
%           figure 3
%==============================
%-------------------------------------------
%                mod004b (late)
%-------------------------------------------
\begin{figure*}
  \begin{center}
     \resizebox{0.9\hsize}{!}{
     \includegraphics[angle=270]{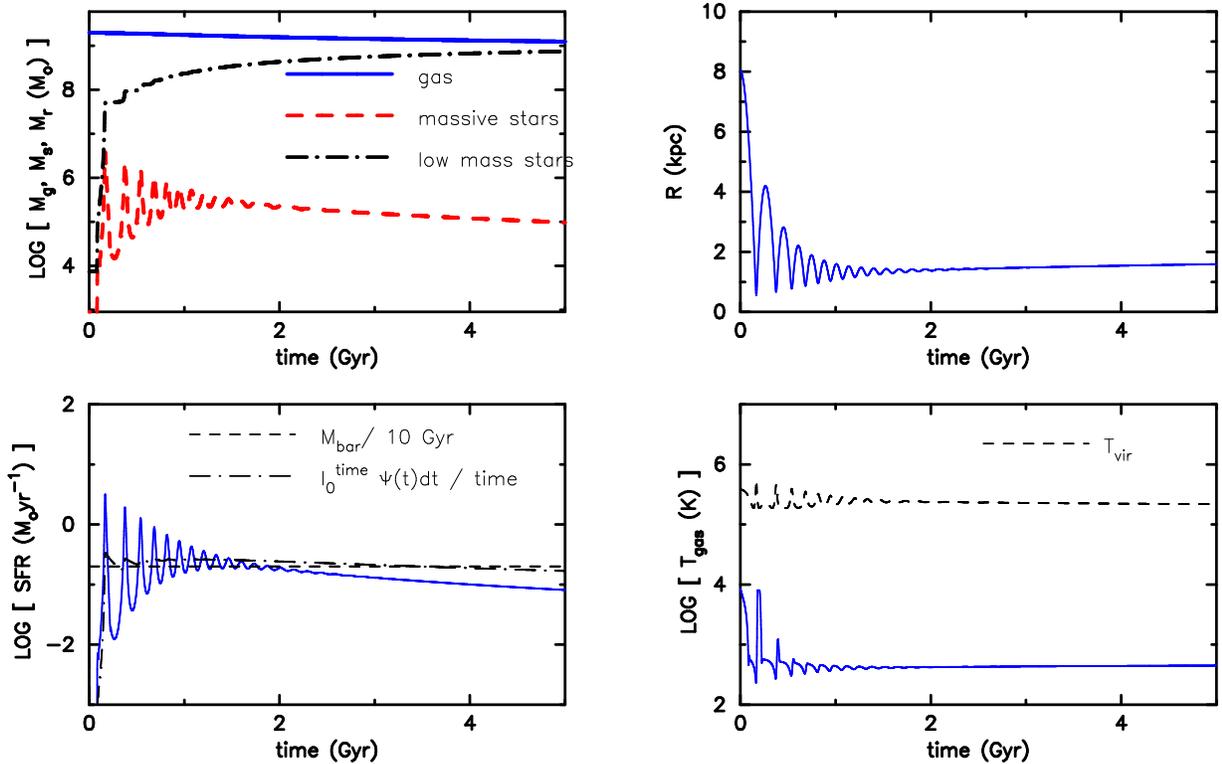}
      }
  \end{center}
  \caption{Same as Fig.\ \ref{fig_ref}, but for 5 Gyr of evolution.}
  \label{theisfig3}
\end{figure*}

   As a first example we present a model with a total baryonic mass of $2
\cdot 10^9 {\rm M}_\odot$ starting at a radius of $R_{\rm ini}=8 {\rm kpc}$.  
The stellar birth function $\Psi_b$ is parametrised by a
Schmidt-exponent of $n=1.5$ and a constant $C_{1.5}=0.06$.
Additionally, a feedback term with $T_s = 100 {\rm K}$ is considered.
The initial temperature is set to $10^5 {\rm K}$ close to the
system's virial temperature
\begin{equation}
    \label{eq_tvir}
    T_{\rm vir} \equiv \frac{\mu m_p}{C_p k} \cdot 
                       \frac{G \left( M_{\rm DM}(R) + \frac{1}{2} M_b \right)}{R}
    \, .
\end{equation}
This definition is motivated by Eq.\ (\ref{d2rdt2smd}) assuming force
equilibrium between thermal pressure at a temperature $T_{\rm vir}$
and the gravitational forces of a nonrotating system, and  $M_{\rm
  DM}(R)$ is the mass of the dark matter component enclosed within the
half-mass radius $R$ of the baryonic component.  The heating rate is
set to $10^{51} {\rm erg}$ per supernova, i.e.\ $h=2.4 \cdot 10^5 \,\,
{\rm pc}^2 {\rm Myr}^{-3}$.

Figure \ref{fig_ref} shows the early evolution: Due to strong cooling,
the gas temperature drops almost immediately to values near $10^4 {\rm
  K}$ bringing the system far out of virial equilibrium.  In the
ongoing collapse, the density increases. Cooling and density
enhancement continue, until star formation becomes more prominent
after about 100 Myr. The first stars reinject energy to the ISM
leading to the quasi-equilibrium stage known from the box models of
KTH95.  Different to them, the density increases because of the continuing
collapse: the temperature evolves on the line of equilibrium
temperatures characteristic of the densities set by the dynamical
state of the system. Though the star formation and, by this, the
stellar energy injection grow, too, the feedback cannot prevent
further collapse, because cooling is still too efficient to allow
the gas to reach the virial temperature. The collapse proceeds
until $t= 0.16 \, {\rm Gyr}$, when the angular momentum conservation
in (\ref{d2rdt2smd}) becomes dominant and the system expands again.  By
this, the first starburst episode comes to an end and the cooling
rate drops.  The gas is quickly heated to $10^4 {\rm K}$;
however, the gas never reaches the virial temperature thanks to the large
increase of the cooling rate at temperatures beyond $10^4 {\rm K}$.
The next star formation cycle starts when the gas is collapsing again.

The longterm evolution of this model shows that the previously
discussed bursts are just transitory phenomena caused by the initial
departure from dynamical equilibrium (Fig.\ \ref{theisfig3}).  The
oscillations are almost completely damped after 2 Gyr. However, only
the first two or three peaks in the SFR are
enough strong to be characterised as a starburst by standard
definitions. Later on no bursts are found.  Such behaviour, i.e.\
only an initial burst, was found in almost all models.

Two general properties of our models can already be seen in the
reference model. First, despite the coupling of stars and gas in the
dynamical equation, Eq.\ (\ref{d2rdt2smd}), the dynamics do (almost)
not care about the star formation and the related stellar feedback.
This is caused by the very efficient cooling resulting in gas
temperatures well below the virial temperature. Second, the SFR is
strongly modulated by the dynamics. Exceptions to both ´´rules´´
exist, but these are either singular or in physically uninteresting
regimes (and beyond the validity of our model equations).

%The evolution
%is only changing, when the mass is reduced to about $10^6 \, {\rm M}_\odot$.
%In that mass regime (or below) the virial temperature is
%of the order of $10^4 \, {\rm K}$. Therefore, the heating due to stellar
%feedback can result in gas temperatures which allow to overcome the
%gravity of the system, i.e.\ a blow-out or a blow-away might take
%place. It is remarkable to note that the overall star formation
%efficiency reaches then at maximum only a few percent.

%--------------------------------------------------
%       Influence of stellar birth function
%--------------------------------------------------

\subsection{Influence of stellar birth function}
\label{sect_res_sbf}

%==============================
%           figure 4
%==============================
%----------------------------------------------------------------
%          SFR (n=1, 1.1, 1.5, 2.0, 2.5, C const, no FB)
%----------------------------------------------------------------
\begin{figure}
  \begin{center}
     \resizebox{\hsize}{!}{
     \includegraphics[angle=270]{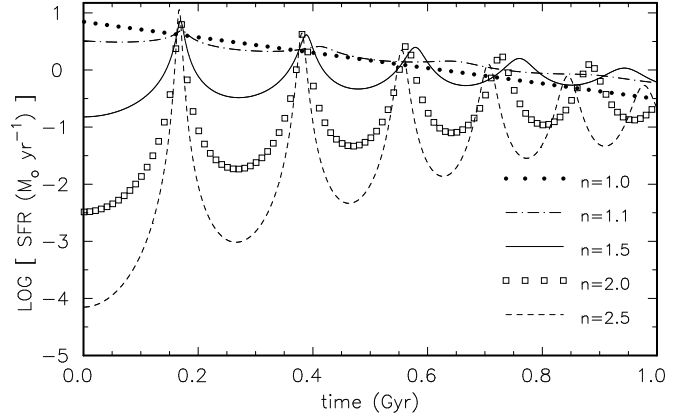}
      }
  \end{center}
  \caption{Temporal evolution of the SFR for models
    with different exponent in the stellar birth function: $n=1$
    (model DN1, filled circles), $n=1.1$ (DN2, dot-dashed), $n=1.5$
    (DN3, solid), $n=2$ (DN4, open squares) and $n=2.5$ (DN5, dashed).
    The normalisation constant $C_n$ is kept constant at 0.007.  No
    negative thermal feedback term is included in the stellar birth
    function.}
  \label{fig_sfr_nvariation}
\end{figure}

%==============================
%           figure 5
%==============================
%-----------------------------------------
%          mod008 (n=1, no FB)
%-----------------------------------------
\begin{figure*}
  \begin{center}
     \resizebox{0.9\hsize}{!}{
     \includegraphics[angle=270]{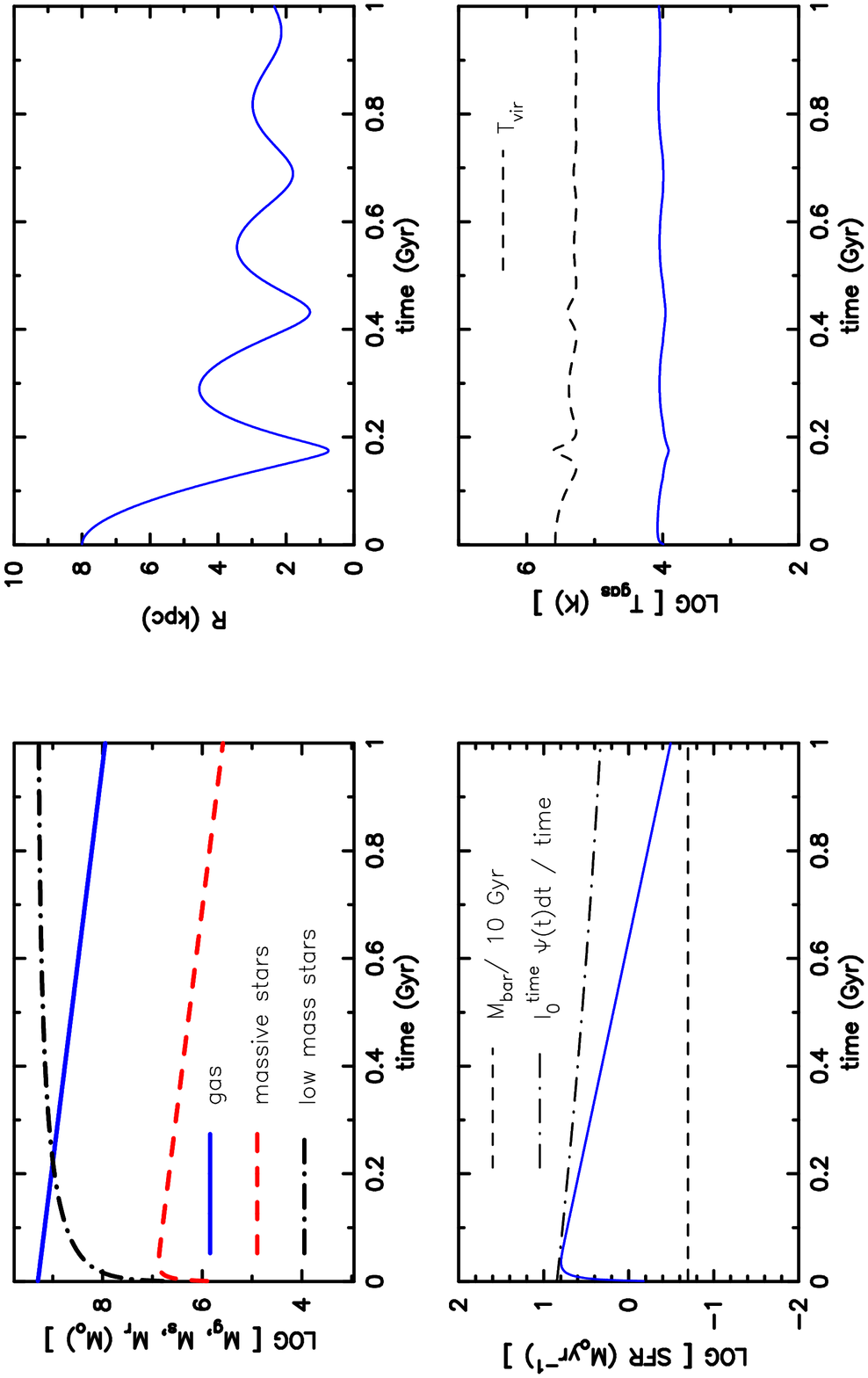}
      }
  \end{center}
  \caption{Temporal evolution of model DN1 characterised by a linear
    Schmidt law ($n=1, C_1=0.007$), but without a negative thermal
    feedback term in the stellar birth function.  For further details
    see Fig.\ \ref{fig_ref}.}
  \label{fig_mod008}
\end{figure*}

%==============================
%           figure 6
%==============================
%-----------------------------------------
%          mod009 (n=1.1, no FB)
%-----------------------------------------
\begin{figure*}
  \begin{center}
     \resizebox{0.9\hsize}{!}{
     \includegraphics[angle=270]{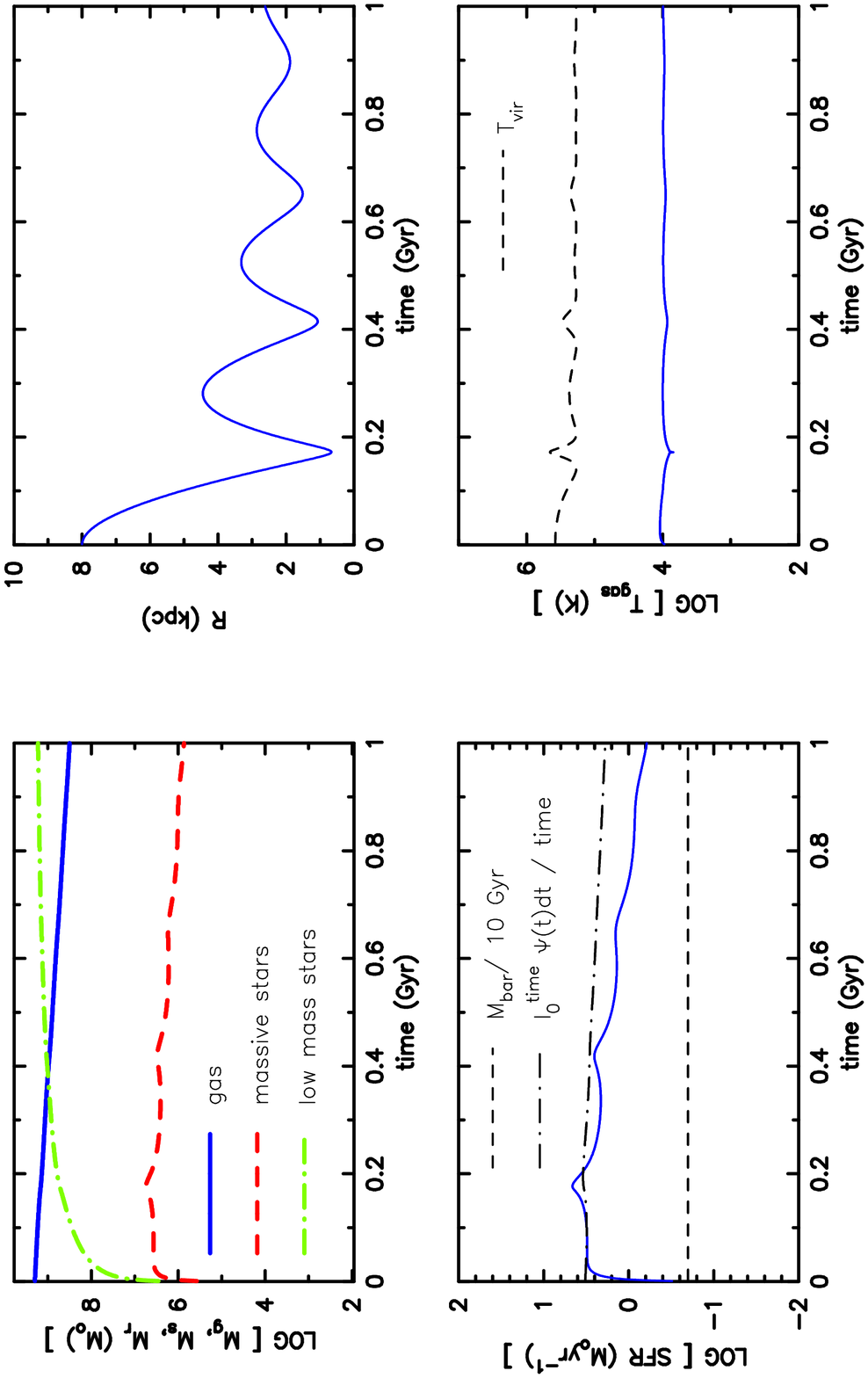}
      }
  \end{center}
  \caption{Temporal evolution of model DN2 characterised by a Schmidt
    law ($n=1.1, C_n=0.007$), but without a negative thermal feedback
    term in the stellar birth function.  For further details see Fig.\
    \ref{fig_ref}.}
  \label{fig_mod009}
\end{figure*}

%==============================
%           figure 7
%==============================
%-----------------------------------------
%          mod011 (n=2, no FB)
%-----------------------------------------
\begin{figure*}
  \begin{center}
     \resizebox{0.9\hsize}{!}{
     \includegraphics[angle=270]{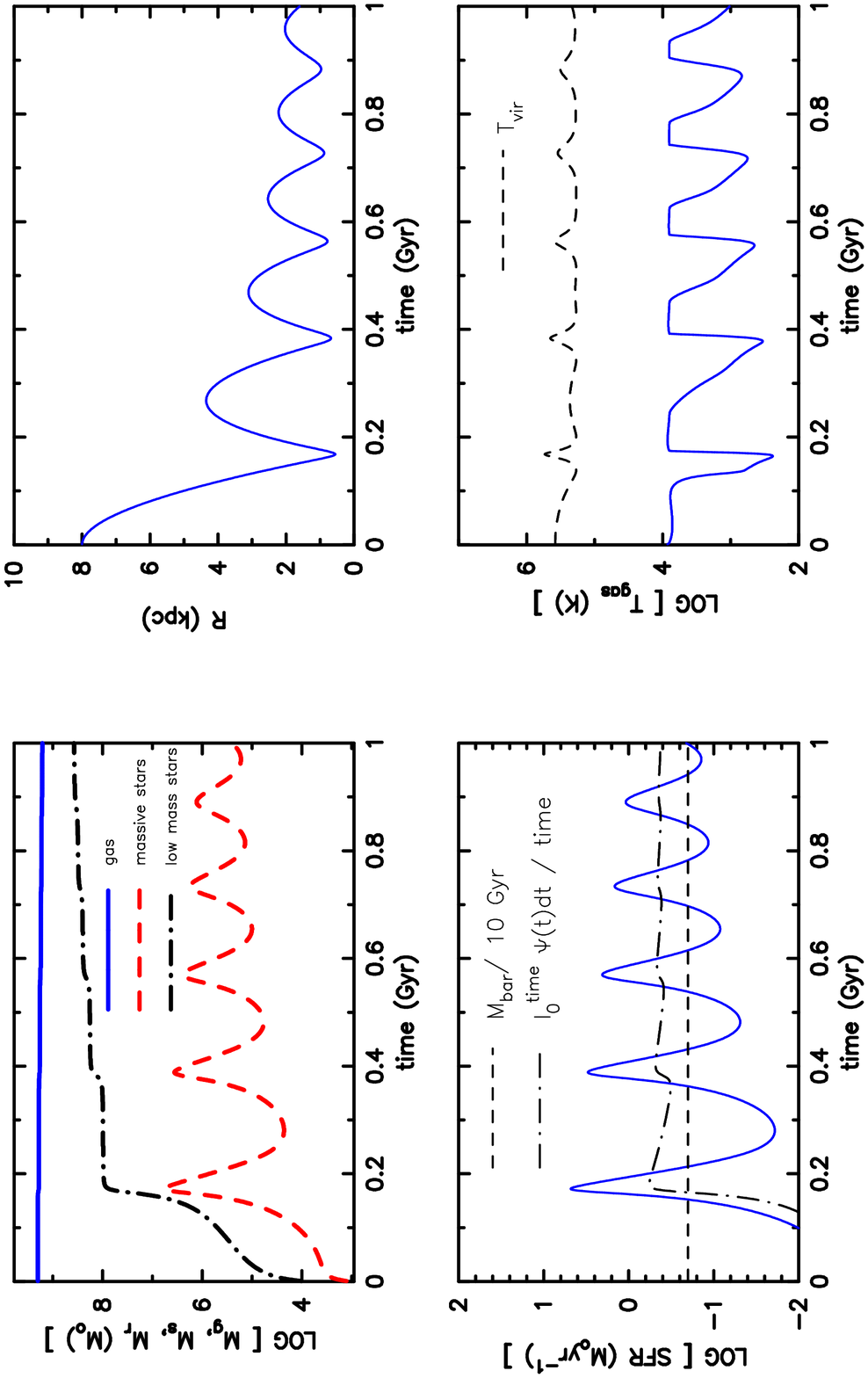}
      }
  \end{center}
  \caption{Temporal evolution of model DN4 characterized by a
    Schmidt law ($n=2, C_n=0.007$), but without a negative
    thermal feedback term in the stellar birth function. 
   For further details see Fig.\ \ref{fig_ref}.}
  \label{fig_mod011}
\end{figure*}

In this section we address 
the influence of the
stellar birth function and its parametrization according to Eq.\
(\ref{psism1}). In a first series of models (DN1 to DN5), we varied the
exponent $n$ in the stellar birth function fixing the constant $C_n$ to 0.007.  We also
neglected the negative thermal feedback; i.e., we consider the limit
$T_s \rightarrow \infty$. This represents the classical Schmidt-law
for which the stellar birth function only depends on density, but not on the temperature of
the gas.

\noindent
{\bf The Schmidt exponent.}
Figure \ref{fig_sfr_nvariation} shows the temporal evolution of the
SFR for different exponents $n$. With increasing $n$,
the SFR becomes increasingly sensitive to the density variations.  The
peaks of the SFR are fairly similar compared to the large SFR
variations, whereas the amount of available fuel varies strongly with
$n$ as a time integration of the SFR shows: for the linear Schmidt law,
it takes 200 Myr to convert 50\% of the gas into stars, whereas for
$n=1.5$ this takes about 1.4 Gyr.  With increasing nonlinearity of
the stellar birth function $\Psi_b$, the terms containing the gas density become more
important for the SFR and the coupling to the dynamics is reflected in
the global SFR. For $n=2$ all oscillations of the SFR within the first
$10^9\, \mathrm{yr}$ have amplitudes greater than one order of
magnitude.

A linear Schmidt law ($n=1$) with no thermal feedback term in the stellar
birth function
represents a singular case with respect to the SFR
$\Psi$. For $n=1$, the SFR is proportional to the total gas mass, but
independent on the gas density: $\Psi = \Psi_b \cdot V = C_1 g \cdot V
= C_1 M_g$. This leads to an exponential decay of the gas mass,
irrespective of the dynamical state of the system: the virial
oscillations do not show up in the exponentially decaying SFR or the
amount of massive stars (Fig.\ \ref{fig_mod008}, left panels).

During the early
evolution, the gas temperature is always close to $10^4$ K
(Fig.\ \ref{fig_mod008}, lower right panel). The high
temperatures are a result of stellar heating. Because of a lack of 
negative thermal feedback in model DN1, the temperature does not
regulate the stellar birth function. Only the strong rise in the cooling
function beyond $10^4$ K prevents the gas from reaching higher
temperatures. On the other hand, the temperatures are too low in the
early stage to become dynamically important as a comparison with the
virial temperatures shows (lower right panel in Fig.\ 
\ref{fig_mod008}). This lack of star formation induced feedback to the dynamics
has already been found in the reference model.
Thus, for the linear Schmidt law, both 
the dynamics and the star formation history are practically decoupled.

\begin{figure}
  \begin{center}
     \resizebox{\hsize}{!}{
     \includegraphics[angle=270]{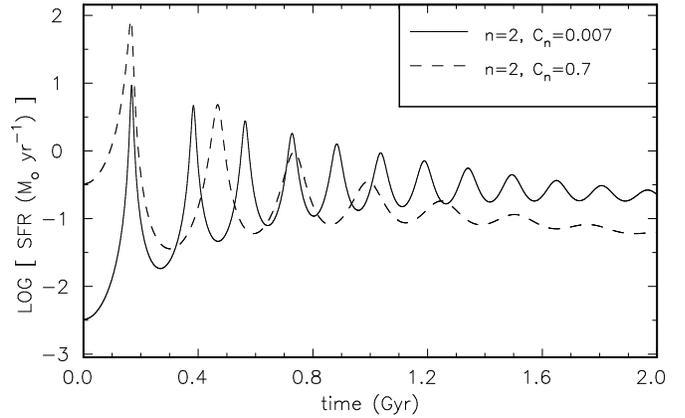}
      }
  \end{center}
  \caption{Temporal evolution of the SFR for the
    models DN4 and DN6 both characterized by a quadratic
    Schmidt law with different normalisation constants:
    $C_n=0.007$ (DN4) and $C_n=0.7$ (DN6).}
  \label{fig_sfr_mod011_mod014}
\end{figure}

\noindent
{\bf The normalisation constant.}
  In a next step we study the influence of the constant $C_n$ 
in the stellar birth function. For a quadratic Schmidt law
we varied $C_n$ by two orders of magnitude.
Initially, the difference in the normalisation constants is directly
reflected in the enhanced SFR of model DN6 (Fig.\ \ref{fig_sfr_mod011_mod014}). 
The enhanced SFR leads to a stronger stellar feedback for model DN6.
By this, the mean radius increases by about 50\% and the period of the variations
is accordingly longer compared to model DN4. The oscillations are more strongly damped 
and already after about 500 Myr the mean SFR is lower for the
model with the larger star formation constant. The latter is a direct consequence
of the lack of fuel after the first strong starburst for
model DN6.

%==============================
%           figure 8
%==============================
%---------------------------------------------------
%     SFR: mod014/011 (n=2, C_n=0.007 and C_n=0.7)
%---------------------------------------------------

\begin{figure}
  \begin{center}
     \resizebox{\hsize}{!}{
     \includegraphics[angle=270]{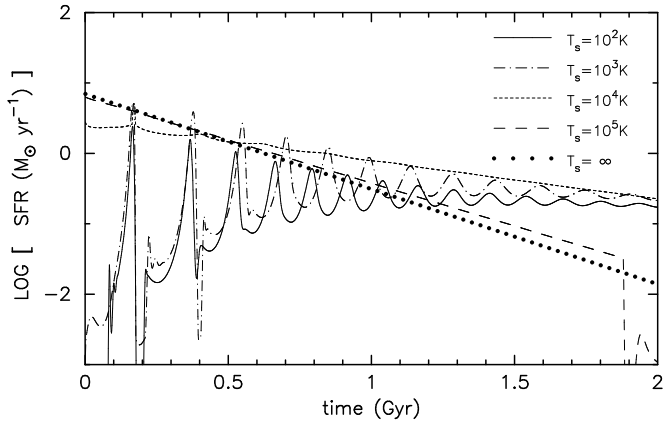}
      }
  \end{center}
  \caption{Temporal evolution of the SFR for models
    with a linear Schmidt law ($n=1$) and a negative thermal feedback
    term with different values of the temperature parameter $T_s$:
    $T_s=\infty$ (model DN1, no thermal feedback in the stellar birth function, filled
    circles), $T_s=10^2\,{\rm K}$ (DT4, solid), $T_s=10^3\,{\rm K}$
    (DT3, dot-dashed), $T_s=10^4\,{\rm K}$ (DT2, short-dashed), and
    $T_s=10^5\,{\rm K}$ (DT1, long-dashed).  The normalisation
    constant $C_n$ is kept constant at 0.007.}
  \label{fig_sfr_tsvariation}
\end{figure}

\noindent
{\bf The negative thermal feedback.} 
So far, we have neglected the thermal feedback $\exp(-T/T_s)$ in the stellar
birth function by
setting $T_s = \infty$. When we consider high, but finite, temperature
scales $T_s$, the SFR evolves similar to the model without any
feedback. Figure \ref{fig_sfr_tsvariation} compares different $T_s$
values for a linear Schmidt law.  In the case of $T_s=10^5 \, \mathrm{K}$
(model DT1) the SFR is almost identic to the case of
$T_s=\infty$ (model DN1). After about 1.9 Gyr the SFR suddenly ceases
caused by the rapid expansion of the gaseous system when the density
drops below a critical density. As discussed in Sect.\
\ref{sect_res_low_gas_densities}, the heating then overcomes the high
cooling rates beyond $10^4$ K and the temperature rises to values
above $T_s=10^5$ K making the feedback term operational.

Reducing $T_s$ to $10^4$ K (model DT2) brings $T_s$ closer to the
actual temperatures of the ISM mainly set by the strong rise in the
cooling function.  Variations in the temperature influence the stellar
birth function much more strongly than in model DT1. This can be seen in the
reduced overall SFR, which shows a modulated
exponential decay. Setting $T_s$ to $10^2$ or $10^3$ K makes the
density variations and its impact on the temperature by cooling to the
dominant factor for the global SFR.  The oscillation
periods are then given by the dynamical timescale and the star
formation reaches a quasi-equilibrium state when the virial
oscillations have been decayed.

%==============================
%           figure 9
%==============================
%--------------------------------------------------------------------
%     SFR: (mod008/040/041/042/043) (n=1, C_n=0.007, T_s varied)
%--------------------------------------------------------------------

\begin{figure}
  \begin{center}
     \resizebox{\hsize}{!}{
     \includegraphics[angle=0]{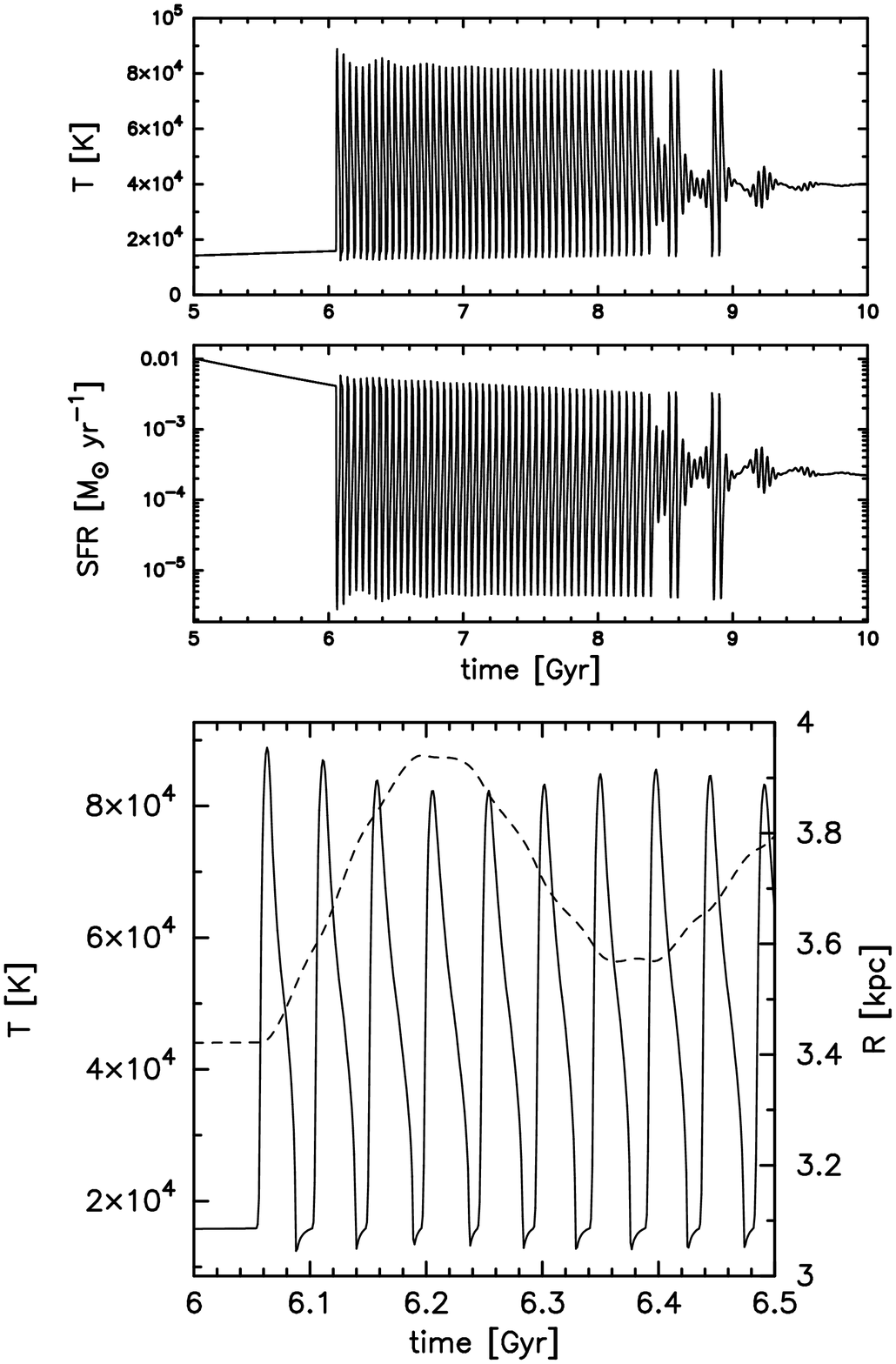}
      }
  \end{center}
  \caption{Late temporal evolution of the gas temperature (upper and
    lower panel, solid line), the SFR (middle panel)
    and the mean radius of the system (lower panel, dashed line) for
    the model with a linear Schmidt law ($n=1$) and a negative thermal
    feedback term with $T_s$ (DT2). The lower panel shows only a short
    fraction of the time displayed in the upper panels. 
%It is clearly
%    visible that the radial oscillations are small and decoupled from
%    the temperature variations.
    }
  \label{fig_tg_mod042}
\end{figure}

It is also interesting to consider the long-term evolution of model
DT2 ($T_s = 10^4\,{\rm K}$, Fig.\ \ref{fig_tg_mod042}): after about 6
Gyr of self-regulated quasi-equilibrium evolution, the SFR enters an
oscillating period over the next 3 Gyr. Then again the evolution
becomes rather stable.  This variation in the SFR is directly caused
by the temperature fluctuations between about $1.4 \cdot 10^4 \,
\mathrm{K}$ and $8 \cdot 10^4 \, \mathrm{K}$ as the upper two panels
in Fig.\ \ref{fig_tg_mod042} show. Since $T_s$ amounts to $10^4 \,
\mathrm{K}$, the feedback term varies by about a factor of $10^3$.  A
comparison of the radial variations and the temperature
variations shows that the fluctations are not caused by the dynamical
evolution (Fig.\ \ref{fig_tg_mod042}, lower panel): the radial
oscillations are too small in amplitude and are not in phase with
the temperature variations.

Though e.g.\ the temperature variations are too large to allow for a
complete explanation by linear perturbation theory, a linear stability
analysis might give a first clue to the existence of this unstable
regime. Since the dynamical evolution is unimportant here, we can
apply the results of the stability analysis for the nondynamical set
of equations (\ref{dgdtsm1}) - (\ref{dedtsm1}) given in KTH95: a
necessary condition for stability is that Eq.\ (28) of KTH95 holds.
In the limit of low gas densities, we can rewrite this stability
criterion as
\begin{equation}
%  - \frac{1/\tau} - \frac{\Lambda^'}(T) \, {\tau_\mathrm{cool}}< 0
    \tau_\mathrm{cool} > - \Lambda'(T) \,\, \tau \, .
   \label{stabcond}
\end{equation}
%with the cooling timescale 
%\begin{equation}
%   \tau_\mathrm{cool} \equiv \frac{e}{g^2 \Lambda} = \frac{bT}{g \Lambda} \, .
%\end{equation} 
This is the case, if the logarithmic slope $\Lambda'$ of the cooling
function is positive definite (as assumed for the whole temperature
regime considered by KTH95). However, the more realistic cooling function 
applied here drops for
temperatures beyond the peaks caused by hydrogen and helium line
emission.  Though this behaviour depends on metallicity (becoming 
weaker with increasing metal content), it persists for metallicities
ranging from metal-free to solar metallicity (B\"ohringer \& Hensler
\cite{boehringer89}). Thus, the stability of the system is not
guaranteed anymore.

Assuming that the logarithmic slope of the cooling function is close to
one, instability can only occur if the cooling timescale is
longer than the mean stellar lifetime $\tau$ of the massive stars. In
the early stages when the gas density is high, the instability
criterion will not be met due to the high efficiency of cooling.
However, in the late stages when most gas is consumed (or lost), the
cooling timescale becomes long even for temperatures above $10^4$ K.

For a metallicity of 1/10 $Z_{\sun}$, the cooling curve according to
B\"ohringer \& Hensler has a first region of negative slope between
$1.5 \cdot 10^4$ K and $4 \cdot 10^4$ K. Therefore, the evolution
becomes unstable once the equilibrium temperature
rises to $1.5 \cdot 10^4$ K. The next unstable region caused
by He line emission exists for equilibrium temperatures close to
$10^5$ K. However, in model DT2 this regime is not reached within the
first 10 Gyr.

%==============================
%           figure 10
%==============================
%-------------------------------------------------------------------------
%     long-term Tg, SFR, R for model 42  (with feedback, T_s=10^4K, DT2)
%-------------------------------------------------------------------------

%--------------------------------------------------
%       Influence of the ISM heating
%--------------------------------------------------

\subsection{Influence of positive stellar feedback}
\label{sect_res_posfeedback}

%==============================
%           figure 11
%==============================
%--------------------------------------------------------------
%       mod004b + 50,52,54,56  (induced & spontaneous SF)
%--------------------------------------------------------------
\begin{figure}
  \begin{center}
    \resizebox{0.9\hsize}{!}{
    \includegraphics[angle=0]{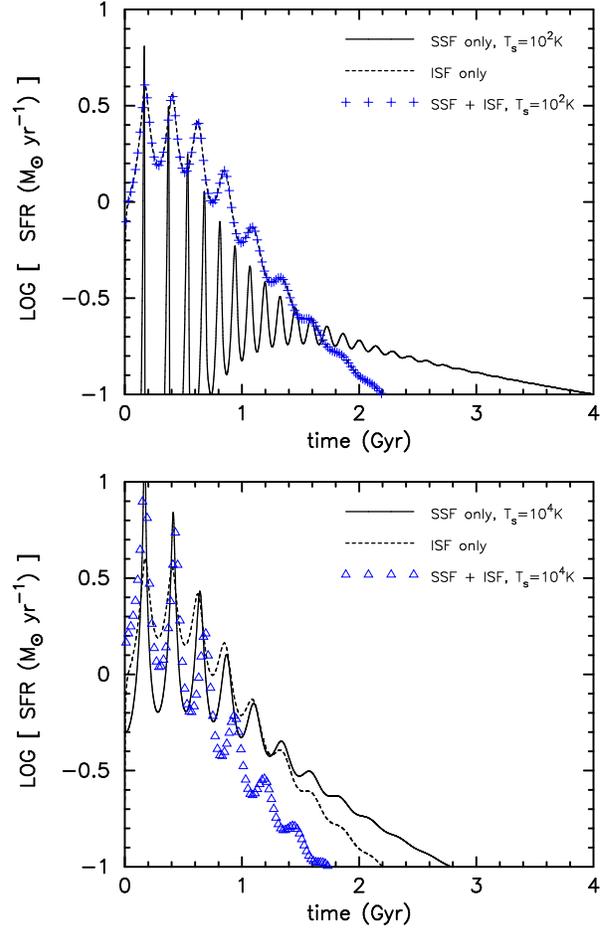}
     }
  \end{center}
  \caption{Temporal evolution of the SFR for models
    with different stellar birth modes: induced star formation only
    (dashed line, DI1), spontaneous star formation only (solid line;
    $T_s$ = 100 K (model A, upper panel) and $T_s = 10^4$ K (A2, lower
    panel)) and both star formation modes together ($T_s$ = 100 K (DI2, plus signs
    in upper panel) and $T_s = 10^4$ K (DI3, open triangles in lower
    panel)).}
  \label{fig_sfr_sfmodes}
\end{figure}

%==============================
%           figure 12
%==============================
%--------------------------------------------------------------
%       mod052 (induced + spont. SF): fractions...
%--------------------------------------------------------------
\begin{figure}
  \begin{center}
    \resizebox{0.9\hsize}{!}{
    \includegraphics[angle=270]{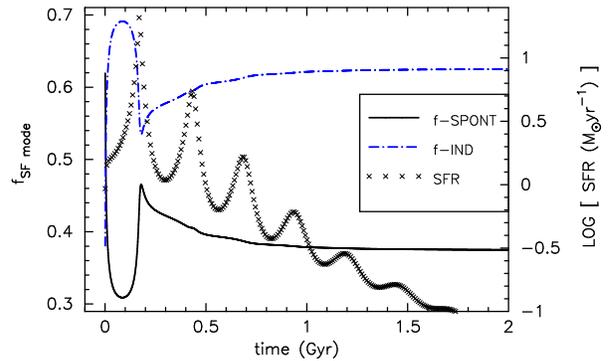}
     }
  \end{center}
  \caption{Temporal evolution of the relative contribution of the
    spontaneous (solid) and the induced (dot-dashed) star formation
    mode to the total mass converted into stars for model DI3.  For
    comparison the combined SFR is also shown (plus).}
  \label{fig_fsfr_sfmodes}
\end{figure}

The star formation recipe discussed in Sect.\ \ref{sect_res_sbf}
corresponds to a spontaneous star formation mode with a negative
stellar feedback. However, there might also be positive stellar
feedback (see e.g.\ the stochastically self-propagating star formation
models (SSPSF) by Gerola \& Seiden \cite{gerola78}).  A model often
discussed is the SN-induced star formation (e.g.\ Ehlerov\'a et al.\
\cite{ehlerova97}, Hosokawa \& Inutsuka \cite{hosokawa05,
  hosokawa06}): the basic idea is that expanding SN-shells sweep up
the ambient ISM in a thin, very dense shell. Such shells cool very
efficiently which results in fragmentation and subsequent star
formation.

To investigate the influence of an induced star formation
mode, we extend our stellar birth function to
\begin{equation}
   \label{psism2}
   \Psi_b(g,T;s,R) \equiv \Psi_{b,{\rm sp}}(g,T) + \Psi_{b,{\rm in}}(g,s,R)  \, ,
\end{equation}
which allows for both a spontaneous star formation mode and an
induced star formation mode.  For the spontaneous star formation mode
$\Psi_{b,{\rm sp}}(g,T)$ we adopt the description applied in the
previous sections; i.e., Eq.\ (\ref{psism1}). For the induced star
formation rate we estimate the gas mass swept up by SN shells and
turned into stars by
\begin{equation}
   \label{psism3}
   \Psi_{b,{\rm in}}(g,s,R) \equiv \frac{\eta_i g}{\tau_i} \cdot f_i(R_{\rm sh}(s,g), R)
   \,\, ,
\end{equation}
where $f_i$ denotes the fraction of a galaxy covered by SN shells, i.e.\ $g
f_i$ is the maximum amount of gas undergoing SN shell-induced star
formation. A fraction $\eta_i$ of this mass is then converted on a
timescale $\tau_i$ into stars.  The factor $f_i$ is related to the
volume $R_{\rm sh}^3$ filled by massive superbubbles undergoing
fragmentation relative to the volume of the galaxy. For simplicity we
assume
\begin{equation}
   \label{psism4}
   f_i(R_{\rm sh}(s,g), R) \equiv 1 - \displaystyle e^{- \left( R_{\rm sh} / R\right)^3}.
\end{equation}
In the case of low star formation activity, only a small fraction of the
galactic volume is affected by induced star formation, whereas for
high star formation activity most of the gas mass might be swept up.
The related star formation timescale ${\tau_i}$ is given by the time
for bubble creation. The efficieny factor $\eta_i$ gives the fraction
of gas accumulated in shells that is converted into stars.  We used
a value of $\eta_i=0.1$ here. The size and the timescale for the
bubble evolution were taken from Eqs.\ (13) and (14) of Ehlerov\'a et
al.\ (\cite{ehlerova97}). It should be noted that this star formation
recipe is only meant to be a simple qualitative estimate, not a
detailed sophisticated induced star formation model. Anyway, we
believe that this ansatz covers all essential ingredients for the
SN-induced star formation mode.

Figure \ref{fig_sfr_sfmodes} displays the SFR for
different combinations of the star formation modes. In case of a low temperature
$T_s = 100$ K, the overall SFR is dominated by the induced star
formation (ISF) mode (upper panel of Fig.\ \ref{fig_sfr_sfmodes}) as
the agreement between model DI1 (ISF only mode) and the model DI2
(combined star formation modes) demonstrates. This is caused by the high
temperatures of about $10^4$ K created by the massive stars formed by
induced star formation. Due to negative thermal feedback the spontaneous star formation is
almost completely suppressed. This suppression can, of course, not
take place in a purely spontaneous star formation model (model A).  In that case,
the amplitudes of the density driven SFR variations increase because of
the temperature dependent variations in the feedback term.  The
dominance of the dynamical evolution is reflected in the almost
identical periods of the SFR in all models. The slightly longer
periods for the models DI1 and DI2 (which include induced star
formation) stem from the larger equilibrium radii leading to a longer
free-fall time.

To have a substantial contribution from the spontaneous star formation mode
in a stellar birth function including induced star formation, the high
efficiency of the thermal feedback has to be reduced. This can be
realized by an increase in the feedback temperature scale $T_s$ close
to or above the actual gas temperature. Model DI3 has a $T_s$ of
$10^4$ K: models using either of the two star formation modes (induced SF,
spontaneous SF) or both of them become then comparable in star formation
amplitudes, peak values and the overall production rate of stars
(Fig.\ \ref{fig_sfr_sfmodes}, lower panel). Still, the SFR
is dominated by the dynamically driven oscillations. It is
interesting to note that after 1 Gyr the combined SFR (model DI3)
becomes smaller than both single-mode models (DI1 and A2).

The relative fraction of the spontaneous and the induced star
formation mode in the combined model DI3 with respect to the total
number of stars formed is shown in Fig.\ \ref{fig_fsfr_sfmodes}: at
the very beginning the spontaneous star formation exceeds the induced
star formation. However, with an increasing overall SFR the induced star formation 
begins to
dominate and the spontaneous star formation drops to 30\%. The contribution
of the induced star formation drops already {\it before} reaching the
first peak of the SFR: this is caused by the dynamics, i.e.\ the
increase in the gas density during the collapse of the system and the
subsequent enhanced spontaneous star formation. The temperature of the gas remains
almost constant close to $10^4$ K during this evolution. After a
Hubble time about 62\% of the stars are produced by induced star
formation.

{\it It is remarkable that adding induced star formation does not
(necessarily) result in a quick consumption of all available gas}.  For
our fiducial model, roughly half of the stars are formed in one
of the two star formation modes. Though this fraction can be varied by
e.g.\ amplifying the thermal feedback term of the spontaneous star
formation mode or changing the efficiency factor $\eta_i$ of the
induced star formation, none of the star formation modes leads to a
completely different behaviour of the system: still the overall star
formation rate is controlled by the global dynamics.

%--------------------------------------------------
%       Influence of the ISM heating
%--------------------------------------------------

%\subsection{Influence of the ISM heating}
%\label{sect_res_ismheating}

%  {\bf XXX do a comparison for different heating 
%         descriptions, i.e.\ $h$ values, $g$-dependence,
%         IMF impact XXX}

%--------------------------------------------------
%       Influence of the IMF
%--------------------------------------------------

\subsection{Influence of the IMF}
\label{sect_res_imf}

%==============================
%           figure 13
%==============================
%-------------------------------------------------------------------------
%     SFR for SF dependent IMF (WK model) and xi variations
%-------------------------------------------------------------------------
\begin{figure}
  \begin{center}
     \resizebox{\hsize}{!}{
     \includegraphics[angle=270]{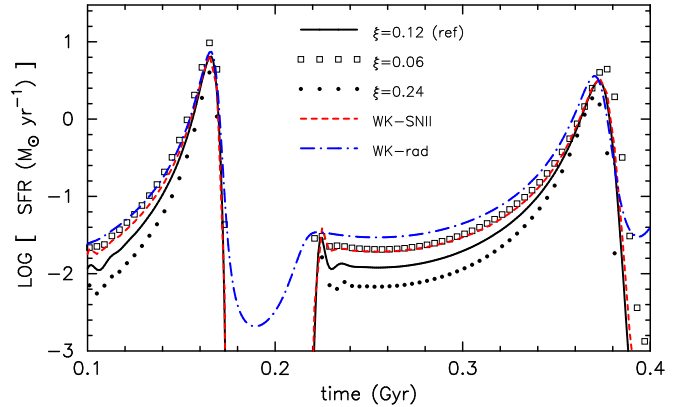}
      }
  \end{center}
  \caption{Temporal evolution of the SFR for the
    reference model A (solid), models DX1 and DX2 with a constant
    IMF-derived fraction $\xi$ of massive stars ($\xi=0.06$: open
    squares; $\xi=0.24$: filled circles) and the Weidner-Kroupa type
    variable IMF models based on SNII heating (DWK1, dashed) and
    radiative heating (DWK2, dot-dashed).}
  \label{fig_SFR_mod045_046}
\end{figure}

{\bf Constant IMF.}  A variation in the stellar initial mass function
will mainly affect the mass fraction $\xi$ of massive stars.
To test limiting cases we varied the time-independent $\xi$ by a
factor of 2. Though such a change is practically beyond the limits of
generally adopted standard IMFs, the impact on the evolution is small;
e.g.\ the SFR is qualitatively identical to the
reference model for both $\xi$-values (Fig.\
\ref{fig_SFR_mod045_046}). The small quantitative differences (e.g.\
lower SFR for larger $\xi$) can be understood as a result of
enhanced ($\xi=0.24$) or reduced thermal feedback. The weak response
of the SFR (and especially its peak values) is a result of the strong
self-regulated coupling of the star-gas system. As a result, the
evolution of the star-gas system is practically insensitive to
reasonable variations in the constant IMF.

\noindent
{\bf Variable IMF.}
Recently, Weidner \& Kroupa (\cite{weidner05}) suggested a variable
IMF that depends on the global star formation activity of the galaxy.
Their main idea is that relations exist between the global
galactic SFR and the maximum mass of the molecular
clouds as well as between the cloud mass and the maximum mass of the
most massive stars in star formation regions.  By this, the global SFR
affects the IMF for massive stars which are
responsible for the main heating and the metal production.

To investigate such a scenario we modified our equations by
introducing an efficiency factor $f_{\rm WK} (\Psi)$ for the
production of massive stars; i.e., we replaced $\xi$ by $\xi \cdot
f_{\rm WK} (\Psi)$.  The efficiency factor was given by a simple
fitting formula adopted from the analysis of K\"oppen et al.\
(\cite{koeppen07}):
\begin{equation}
   f_{\rm WK} = 
  \left\{ 
  \begin{tabular}{ccc}
     {$1 - 0.8 \cdot e^{-x/2}$} &           & {$x \ge 0$} \\
                                & {\rm for} & \\
     {$0.2 \, e^x$}             &           & {$x < 0$} \\
  \end{tabular}
  \right.
\end{equation}
with $ x \equiv 3 + \log \left[ \Psi / (\mathrm{M_\odot \, yr}^{-1})\right]$.  
One should note that the efficiency factor only varies
slightly over the physically interesting regime of
$x>0$ or $\Psi > 10^{-3} \, \mathrm{M_\odot \, yr}^{-1}$.

We considered two cases for the energy feedback. In the first case we
assumed that the feedback is given by type II supernovae (model DWK1).
In that case the heating coefficient corresponding to the energy input
per supernova has to be multiplied by the number of massive stars. The
number of massive stars, however, depends mainly on the lower mass
limit of the mass range of massive stars, and not on the upper mass
region that is mainly affected by a Weidner-Kroupa type IMF. A
detailed calculation shows that the conversion factor between the
number of massive stars and their mass varies only by a factor of 2
when changing the SFR from $10^{-3} \, \mathrm{M_\odot
  \, yr}^{-1}$ to $100 \, \mathrm{M_\odot \, yr}^{-1}$.  In our model
DWK1 we neglect this small variation and keep the energy equation.

In the second case, we assume that the heating is mainly done via
radiation (model DWK2).  Then, the heating coefficient $h$ becomes a
function of the SFR, because of the strong dependence
of the stellar radiative energy input on the upper stellar masses
caused by the mass-luminosity relation. A detailed calculation shows
that the corresponding efficiency factor is similar to the one for the
mass fraction. Therefore, we use for simplicity the same factor
$f_{\rm WK}$, i.e.\ $h$ is replaced by $h \cdot f_{\rm WK}(\Psi)$ in
Eq.\ (\ref{dedtsmd}) (or Eq.\ (\ref{dedtsm1}), respectively).

Figure \ref{fig_SFR_mod045_046} shows the evolution of both models.
The difference between models with and without a variable IMF are
marginal in both, amplitude and timing. Compared to the reference
model A, the Weidner-Kroupa-type IMF models have a slightly enhanced
SFR. This enhancement is necessary to balance the
deficit in the stellar feedback caused by the reduced number of
massive stars in the Weidner-Kroupa-type IMF models.

In general the qualitative evolution of the models is not affected by
the shown IMF variations, either if we change the mass fraction $\xi$
assuming a stationary IMF or if we apply the temporal variable 
star formation-dependent 
IMF suggested by Weidner \& Kroupa (\cite{weidner05}). The main reason
is the very efficient self-regulation.

%--------------------------------------------------
%       Influence of the total baryonic mass
%--------------------------------------------------

%\subsection{Influence of the total baryonic mass}
%\label{sect_res_barmass}

%--------------------------------------------------
%       Influence of dark matter halo
%--------------------------------------------------

%\subsection{Influence of dark matter halo}
%\label{sect_res_darkmass}

%  {\bf XXX mention and compare with McLow \& Ferrara XXX}

%###########################################
%            Discussion
%###########################################

\section{Discussion}
\label{sect_discussion}

%--------------------------------------------------
%       Timescales
%--------------------------------------------------

\subsection{Timescales}
\label{sect_res_timescales}

Basically four physically distinct timescales are involved in our
description: the heating and cooling timescales, $\tau_{\rm fb}$ and
$\tau_{\rm diss}$, the dynamical timescale $\tau_{\rm dyn}$ and the gas
consumption (or star formation) timescale $\tau_{\rm SF}$. KTH95
already has shown that $\tau_ {\rm SF}$ is usually longer than the
timescales governing the energetics of the ISM. This also holds in case
of a multiphase ISM model (K\"oppen et al.\ \cite{koeppen98}), 
where
the timescale for the additional equilibrium between evaporation and 
condensation of the gas clouds is well separated from the other two,
leaving them unperturbed. Therefore, we expect that inclusion
of a multiphase ISM model would not alter the basic behaviour of our
system.

For typical mean gas densities the dynamical timescale exceeds also
the cooling and heating time (though it is less than the gas
consumption time). Therefore, the system almost instantaneously
adjusts to the equilibrium state given by the actual gas density.
Provided the temperature remains in a regime with a positive slope of
the cooling function, the system is then stable. This holds e.g.\ for
temperatures below $10^4$ K, (which is practically the temperature
regime for high and moderate gas densities).  Thus, the radial
oscillations are only initial transient virial oscillations acting on
a dynamical timescale.

This behaviour turned out to be rather robust against variations in
model parameters like stellar heating, gas mass, initial spatial
extension, stellar birth function (with the exception of a linear 
stellar birth function
resulting in an exponentially decaying SFR).  Also
different IMF parametrizations or even a variable IMF (like the
integrated galactic IMF suggested by Weidner \& Kroupa
\cite{weidner05}) do not change the behaviour of the system
qualitatively provided the feedback mechanisms are not cutoff.

%--------------------------------------------------
%       Generic behaviour
%--------------------------------------------------

\subsection{Generic behaviour}
\label{sect_generic_behaviour}

%==============================
%           figure 14
%==============================
%--------------------------------------------------------------
%       mod004b + 6 (phase space)
%--------------------------------------------------------------
\begin{figure}
  \begin{center}
    \resizebox{0.95\hsize}{!}{
    \includegraphics[angle=270]{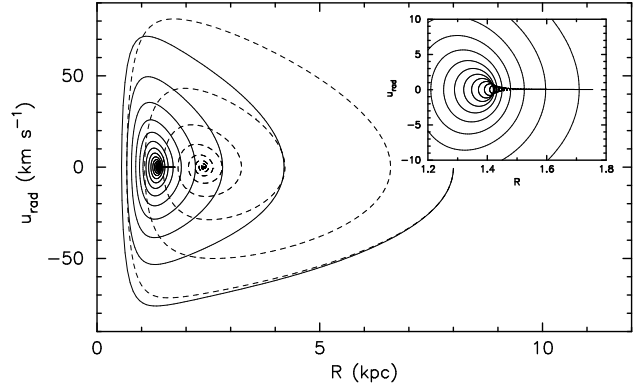}
     }
  \end{center}
  \caption{Evolution of the reference model A (solid) and the model
    DT5 (quadratic Schmidt law with feedback term, $T_s = 10^5$ K;
    dashed line) in the phase space $R$-$v_{\rm rad}$.  The inset
    shows a blow-up of model A near its equilibrium position reached
    after the initial oscillations. 
%A slow drift of the equilibrium
%radius due to gas consumption can be seen.
   }
  \label{fig_phasespace}
\end{figure}

%==============================
%           figure 15
%==============================
%--------------------------------------------------------------
%       mod004b (R-T-plane)
%--------------------------------------------------------------
\begin{figure}
  \begin{center}
    \resizebox{0.95\hsize}{!}{
    \includegraphics[angle=270]{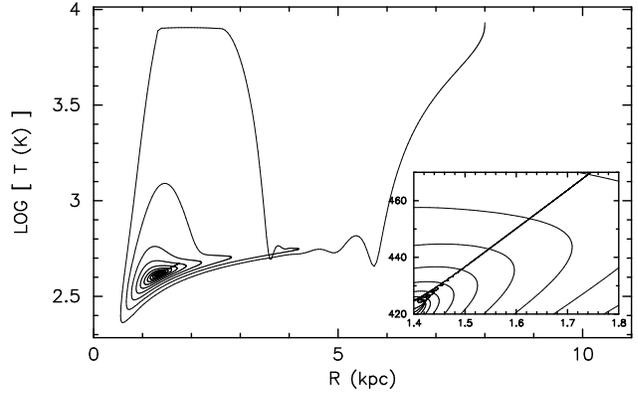}
     }
  \end{center}
  \caption{Evolution of the reference model A in the $R$-$T$ plane.
    The inset shows a blow-up near the equilibrium position reached
    after the initial oscillations. In the inset linear 
    temperature values are used. 
%A slow drift of the equilibrium
% position due to gas consumption can be seen.
    }
  \label{fig_rt}
\end{figure}

Our calculations show that the evolution of the coupled star-gas
system is dominated by the dynamical state of the system. Due to the
short heating and cooling timescales, the ISM relaxes quickly.
Thus, the system's behaviour is governed by the 
dynamical equation Eq.\ (\ref{d2rdt2smd}),
which has the structural form
  \begin{eqnarray}
     {dR\over dt} & = & v_{\rm rad} \label{e:speed} \\
     {dv_{\rm rad}\over dt} & = & -{A_2 \over R^2}
                        + {A_1 \over R} + {A_3 \over R^3}
                        - A_4 v_{\rm rad} 
                                  \label {e:radius} 
  \end{eqnarray}
  with coefficients $A_i$ that are positive for physically sensible
  descriptions.  For simplicity we only consider the self-gravity term
  here.  With a proper choice of the time unit, $A_4$ can be set to
  unity.

  For $v_{\rm rad} = 0$ there exist equilibrium positions, where the
  forces cancel out
  \begin{equation}
     {A_1 \over R_e} - {A_2 \over R_e^2} + {A_3 \over R_e^3} = 0
  \end{equation}
or
  \begin{equation}
     A_1 R_e^2 - A_2 R_e + A_3 = 0
  \end{equation}
with the formal solution
  \begin{equation}
     R_{e\pm} = {1\over 2 A_1} \left( A_2 \pm \sqrt{A_2^2 - 4 A_1 A_3} \, \right)  \, .
  \end{equation}
  If $A_2^2 > 4 A_1 A_3$, two real equilibrium radii exist. Note that
  for $A_1 = 0$ (i.e.\ a gas temperature well below the virial
  temperature) one simply has a single solution
  \begin{equation}
     R_e = {A_3 \over A_2} \, .
  \end{equation}
 
  The analysis of the local stability of these equilibrium points (via the 
  linearized equations) of the Jacobi matrix ${\bf J}$ 
  \begin{eqnarray}
     {\partial \dot{R} \over \partial R} & = & 0 \\
     {\partial \dot{R} \over \partial v_{\rm rad}} & = & 1 \\
     {\partial \dot{v}_{\rm rad} \over \partial R} & = &  -{A_1 \over R_e^2} 
                                   + {2 A_2 \over R_e^3} - {3A_3 \over R_e^4}\\
     {\partial \dot{v}_{\rm rad} \over \partial v_{\rm rad}} & = & = -1 %-A_4
  \end{eqnarray}
  with
  \begin{eqnarray}
     {\rm Trace} \, {\bf J} & = &  - 1 \\ % A_4\\
     {\rm det} \, {\bf J}   & = &  {A_1 \over R_e^2} - {2 A_2 \over R_e^3} 
                                 + {3A_3 \over R_e^4}\\
     {\rm Discr}         & = &  1 - 4 {\rm det} \, {\bf J} /({\rm Trace} \, {\bf J} )^2
  \end{eqnarray}
  shows if equilibrium solutions exist ($A_2^2 > 4 A_1 A_3$) that
  $R_{e-}$ always is an attracting focus, since the trace and the
  discriminant are negative for all positive coefficients $A_k$.
  Likewise, the other point $R_{e+}$
  always is a saddle point (negative determinant).\\

  The consequences of this general structure are shown by a few
  representative trajectories in the phase space. In Fig.\
  \ref{fig_phasespace} we depict these already for the complete system
  of our equations. One notes that if the initial conditions are
  close enough to the attracting focus, the system performs
  damped nonlinear oscillations before settling in the equilibrium.
  If the initial radius or the initial speed is too large, there will
  be a collapse, followed by an expansion. This takes the system to
  the vicinity of the saddle point, whence the expansion continues
  without bounds.

  If one looks more closely at the behaviour in the equilibrium state
  (shown in the inset in Fig.\ \ref{fig_phasespace}), one finds that
  the equilibrium radius does not stay constant, but always increases,
  more or less rapidly depending on the coefficients $A_i$.  The
  reason is that the coefficient $A_1$ of the pressure term in the
  full set of equations depends on the gas temperature. When the
  system reaches the (dynamical) equilibrium, the gas continues to be
  consumed by star formation. The cooling of the gas therefore slows down,
  and the gas temperature increases slowly, giving rise to a slow
  expansion that contributes to the decrease in the gas density.
  This rise in temperature causes an increase in the equilibrium
  radius, which the system simply follows. This leads to the slow
  drift seen in the inset of Fig.\ \ref{fig_phasespace}.

  In Fig.\ \ref{fig_rt} we plot a trajectory in the radius-temperature
  plane. One notes that the oscillations in radius and temperature go
  with a nearly constant phase lag, but once the oscillations have
  damped out, a steady secular increase in both radius and temperature
  takes over (cf.\ to inset in Fig.\ \ref{fig_rt}).

%------------------------------------------------------------
%       Heating-cooling equilibrium at low gas densities
%------------------------------------------------------------

\subsection{Heating-cooling equilibrium at low gas densities}
\label{sect_res_low_gas_densities}

The equilibrium between heating and cooling plays a crucial role for
the overall evolution of our system. Cooling is very efficient for typical 
galactic mean densities, thereby compensating for the
stellar heating processes. However, when the gas density drops, the
heating by stars ($h M_s \propto h \Psi_b V \propto M_g$) can overcome
the cooling ($\Lambda g^2 V \propto g M_g$), so the gas quickly heats
up (here we assume a linear stellar birth function). Once
temperatures become higher than the virial temperature, the gaseous
system expands and a mass loss occurs even without any dramatic event
like a starburst. The critical density $g_{\rm loss}$ for gas loss
can be estimated from the balance between heating and cooling from
Eq.\ (\ref{dedtsm1}) with $s \approx \zeta \Psi \tau = \zeta C_n
g_{\rm loss}^n e^{-T/T_s}$
\begin{equation}
    \label{eq_gloss}
    g_{\rm loss}^{2-n} = \frac{h \zeta C_n \tau e^{-T/T_s}}{\Lambda(T)} \, .
\end{equation}
For the values of model DN1 (linear Schmidt law), we get $g_{\rm loss}
\sim 8 \cdot 10^{-5} \, {\rm M}_{\sun} \, {\rm pc}^{-3}$ by adopting a
cooling rate of \mbox{$\Lambda \approx 10^{-21.5} \, {\rm erg \,\,
    cm}^3\,{\rm s}^{-1}$} characteristic of temperatures well above
the drop near $10^4$ K. The $g_{\rm loss}$ becomes smaller with
increasing Schmidt exponent $n$. Practically, such low gas densities
might not be relevant, e.g., because of the gas replenishment by 
low-mass stars.

It is interesting to note that no critical density exists for a quadratic 
Schmidt law. If $n$ is higher than 2, the system would
even run into a cooling catastrophe.  In the case of optically thin heating,
$h$ in Eq.\ (\ref{eq_gloss}) has to be replaced by $\tilde{h} \, g$.
Thus, the critical exponent for a cooling catastrophe already becomes
$n=1$.  This means that, for all reasonable stellar birth functions, the gas cools down
below a critical density.  In that case other heating mechanisms
providing a steady energy source independent of the amount of massive
stars (hence, the SFR and the gas density) might become significant.

%----------------------------------------------------------
%       Is the dynamics influenced by stellar feedback?
%----------------------------------------------------------

\subsection{Are the dynamics influenced by stellar feedback?}
\label{sect_feedback_dynamics}

%==============================
%           figure 16
%==============================
%--------------------------------------------------------------
%       mod004b (R-T-plane)
%--------------------------------------------------------------
\begin{figure}
  \begin{center}
    \resizebox{0.95\hsize}{!}{
    \includegraphics[angle=270]{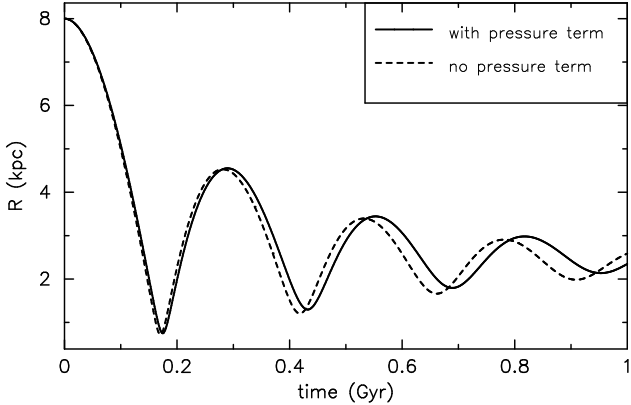}
     }
  \end{center}
  \caption{Temporal evolution of the mean radius of model DN1 (solid)
    and the same model, but neglecting the pressure term in the
    dynamical equation Eq.\ (\ref{d2rdt2smd}) (dashed). }
  \label{fig_r_nopress}
\end{figure}

In this paragraph we discuss the impact of stellar feedback
  on the dynamical evolution.  The latter is
  described by Eq.\ (\ref{d2rdt2smd}) which contains a pressure
  gradient term: $1/g \cdot dP/dr \sim T/R$. Stellar feedback
  can influence the dynamical evolution via this term by heating. Practically speaking,
  this happens when the pressure term is close to or exceeds
  the gravitational acceleration terms (the first two terms in Eq.\
  (\ref{d2rdt2smd})), i.e.\ when the gas temperature is close to
  the virial temperature or higher.
%A major motivation for our analysis was to study the impact of stellar
%feedback (i.e.\ of star formation) on the dynamical evolution of
%(self-)gravitating star-gas system. In principle, Eq.\
%(\ref{d2rdt2smd}) allows for a coupling between dynamics and stellar
%feedback through the pressure term proportional to $1/g \cdot dP/dr
%\sim T/R$.  Practically the dynamics is only influenced by star
%formation, when the gas temperature is of the order of the virial
%temperature or larger.  
However, in case of the evolutionary stages
shown in this paper the gas temperatures are always of the order
$10^4$ K or below, whereas the virial temperature is 
at least a few $10^5$ K.  Physically, the low gas temperatures are
caused by the steep rise in the cooling function beyond $10^4$ K (cf.\
e.g.\ Fig.\ 1 in B\"ohringer \& Hensler \cite{boehringer89}). For
typical gas densities, unrealistically high heating rates are required
to reach (and keep) temperatures beyond $10^4$ K.

Therefore, the dynamics are almost unaffected by the thermal state of
the gas.  This is nicely illustrated in Fig.\ \ref{fig_r_nopress}
where the mean radius of model DN1 is compared with a model neglecting
the pressure term of Eq.\ (\ref{d2rdt2smd}): no substantial difference
between both models is discernible. Small differences develop at
maximum compression when the gravitational terms and the angular
momentum almost cancel. For this short moment, the small pressure-gradient 
term contributes recognisably. It acts as a reduction of the
gravity, e.g.\ like a density reduction, which results in slightly
longer periods.

The situation only changes when either the gas temperature rises or
the virial temperature decreases.  The latter can happen for systems
of lower mass, e.g., for compact systems like globular clusters
(no DM, $\sim 10^6 \mathrm{M}_{\sun}$) or more extended systems like
dSph galaxies, when the virial temperatures drop down to a few $10^4$ K (we
will discuss this in a forthcoming paper). Alternatively, the gas
temperature might be heated to a few $10^5$ K. Practically speaking, this is
only the case where the criterion for run-away heating, Eq.\
(\ref{eq_gloss}), is fulfilled, i.e.\ when the gas density drops below
a critical value due to gas consumption. The actual critical values, however,
are too low, to be practically relevant within the limits of
our model.

The physical reason for the general insensitivity of the
dynamics to the thermal feedback induced by stars is the steep rise in
the cooling function near $10^4$ K. As long as radiative cooling is
the main agent of energy dissipation and as long as the gas
  temperature is kept below $10^4$ K, thermal feedback will only
  affect the dynamics of systems with a virial temperature near $10^4$
  K.  However, for short periods (especially at early stages), the gas
  temperature might reach or exceed temperatures of $10^5$ K briefly affecting
  the dynamics of systems with virial temperatures of that
  order.

However, thermal pressure (and therefore classical cooling) does not need
to be the main agent counter acting gravity. If the system is
composed of sufficiently compact objects, their velocity dispersion
will provide a kinetic pressure, e.g., in case of a cloud system, the
internal temperature of the clouds is irrelevant for their overall
dynamics, which is driven by the velocity dispersion. In that case,
inelastic cloud-cloud collisions provide the relevant energy
dissipation mechanism.  Different from radiative cooling, there is no
steep increase in the cloud-cloud ``cooling'' function and the
arguments preventing a strong mutual coupling between dynamics, star
formation, and stellar feedback are not applicable. Additionally, the
timescale for energy dissipation due to cloud-cloud collisions is much
longer than the radiative cooling timescales, which also might allow
for more of a variety of responses by the system. We will discuss the
related aspects in a future paper.

%--------------------------------------------------------------
%       Comparison with other numerical models
%--------------------------------------------------------------

\subsection{Comparison with ``full'' dwarf galaxy simulations}
\label{sect_comp_num_models}

\noindent
  How do our simple models compare with detailed numerical
  simulations?  Though important, it is a difficult question. At the
  moment there are only a small number of sophisticated simulations
  available.  They differ by many aspects between themselves and from
  our calculations: e.g.\ with respect to the implemented physics, the
  numerical methods or the initial conditions and the general physical
  set-up. Unfortunately, a standard test for dwarf galaxy
  evolution/physics does not exist.

  With all these differences in mind a detailed comparison is not
  meaningful, but a qualitative comparison should shed light on the
  applicability of our ansatz; e.g., Pelupessy et al.\
  (\cite{pelupessy04}, hereafter PWI04) model the evolution of a
  rotating disk-like dwarf galaxy embedded in a dark matter halo. They
  apply a sophisticated ISM model including heating, cooling and an
  ionisation/recombination scheme. Their SFR develops
  a quasi-periodic behaviour with a period of about twice the
  dynamical time for (weak) bursts.  PWI04 interpret this behaviour as
  the result of a kick that drives the gas out of the disk and the
  subsequent fallback of the gas initiating the next star formation episode. This
  concept is very similar to our adopted coupling between feedback and
  dynamics, though we are investigating a spherical system.

  The overall star formation peaks in PWI04 are much smaller than in our
  calculations (by one order for a fiducial model including a feedback
  term). This might be caused by the different initial set-ups, because the
  spherical configuration we study allows for a much deeper collapse
  than a rotationally supported disk. As a result, the density and,
  thus, the star formation variations are much higher in our case than for PWI04.
  For comparison, the full 1d chemo-dynamical models of Hensler et
  al.\ (\cite{hensler04}, afterwards HTG04) model spherical galaxies
  (but without dark matter) and they exhibit SFR
  comparable to ours.  Similar to our calculations, the models by
  HTG04 show a strong trend toward self-regulation that has also been found in
  the 3D chemo-dynamical N-body-SPH models of Berczik et al.\
  (\cite{berczik02}) including dark matter.  The PWI04 models exhibit
  no strong starbursts as the moderate variation in their star
  formation rates shows (Figs.\ 4 and 6 in PWI04).  In contrast to
  them we were looking for strong variations that are rarer, but
  that exist (van Zee \cite{vanzee01}).

  Another interesting example of SFR variability is the SPH simulation
  of Struck (\cite{struck05}, hereafter S05), who investigated the
  evolution of a disk galaxy of about $2 \cdot 10^{11} \, M_\odot$.
  The gaseous component in this simulation was subject to star
  formation, stellar feedback (by SNII) and dissipation (radiative
  cooling).  S05 finds a strongly variable star formation with two
  modes: local, off-centre gas concentrations can lead to local,
  ``premature'' bursts preventing the system from a global starburst.
  The result is an incoherent star formation pattern with an
  irregular, mildly variable SFR.  Sometimes, however,
  large-scale motions coherently funnel matter into the central region
  where a strong starburst is then ignited (cf.\ Figs.\ 1 and A1 in
  S05).  The related timescale is about twice the free-fall timescale
  of a particle at 2-3 kpc above the plane.  This burst mode
  corresponds physically to the oscillatory ``breathing'' mode seen in
  our calculations, though the masses of the models differ by one
  order of magnitude. Similarly, the burst duration (about 10 Myr) is
  comparable to the burst duration in our models.  On the other hand,
  our simple model is unable to resolve purely local effects leading
  to incoherent star formation modes.

  Though a detailed comparison between our simple model and the
  different complex numerical simulations is unreasonable because of
  differences in implemented physics and in initial conditions, the
  qualitative agreement is obvious: the bursts essentially stem from 
  the oscillations of the global dynamics. The gas simply has to
  follow them, so the starbursts could appear in any system. The
  tight coupling between the large-scale dynamics and the occurrence of
  global (strong) starbursts is corroborated in all the different
  studies, by this motivating and strengthening our ansatz.

%--------------------------------------------------------------
%       A speculation about bursts in interacting galaxies 
%--------------------------------------------------------------

\subsection{A speculation about bursts in interacting galaxies...}
\label{sect_res_interactions}

\noindent
By definition our model refers to isolated systems. However, the
initial conditions leading to a collapse of the system could also stem
from galaxy interactions. Therefore, one might speculate about the
possible implications for interacting galaxies.

An important result of our simulations is the strong coupling of the
SFR to the dynamics (with the discussed exception of linear Schmidt
law without feedback term). To create a starburst, the gas
density must be strongly enhanced, otherwise the self-regulation
prevents large SFR variations. The availability of large amounts of
gas alone is not sufficient for a starburst as the example
of poststarburst E+A galaxies with a large amount of HI gas demonstrated
(Buyle et al.\ \cite{buyle06}).

For interacting galaxies, one might speculate that only those
perturbations leading to structural rearrangements, e.g.\ a
substantial inflow of gas or a merger, end up in a
starburst.  Whether galaxies produce sufficient
density enhancements, hence a starburst, depends not only on the
existence of a proper perturbation (like an interaction or an internal bar)
but also on the internal dynamical stability of the involved galaxies.
This might explain the wide variety of star formation responses found in simulations
of interacting galaxies (e.g.\ in Di Matteo et al.\ \cite{dimatteo07}).

%--------------------------------------------------
%        Simplifications of the model
%--------------------------------------------------

\subsection{Simplifications of the model}
\label{sect_res_simplifications}

    Describing galaxy evolution by a few equations always requires
simplifications. Here we want to comment on a few simplifications (or
neglections).

\noindent
{\bf Dynamics.}
The dynamics of the system is summarily described by a mean radius and
its temporal change. A decomposition into several structural
components (like a disk-bulge-halo structure) is beyond the limits of this model.
Moreover, density gradients or even an asphericity are not considered.
An extension of our simple model might be possible. However, with
respect to structure formation, we think that the related physics are
too complex to be condensed reliably into another ``simple'' equation.

Implicitly we also assumed that the dynamics of the different baryonic
components can be modelled by the same mean radius. This need not to
be true, of course.  Strictly speaking this is only valid when the
dynamics of stars and gas is either subject to the same acceleration
terms (e.g.\ similar pressure/velocity dispersion terms) and/or when
the self-gravity of the stars is not important for the dynamical
evolution of the gas. The latter is the case in gas-rich galaxies or
in dark-matter-dominated potentials (when the self-gravity term
remains unimportant). Since we are interested here in starburst
galaxies that are mainly gas-rich dwarf galaxies, the restriction to
gas-dominated stages seems reasonable. The dynamical equations can be
easily extended to several distinct components, if required.

\noindent
{\bf ISM description.}
In our approach we adopted a single ISM phase. An extension to a
multi-phase ISM model like in K\"oppen et al.\ (\cite{koeppen98})
would be possible 
and is envisaged. However, we do not expect significant changes, 
because the timescales related to the additional multi-phase ISM 
processes (like condensation, evaporation) are separated well from 
the other timescales: they are much longer than the dominant heating 
and cooling processes, so that thermal equilibrium is established 
rapidly and independently of the fractional ratio of clouds and 
intercloud gas. On the other hand, evaporation and condensation of 
clouds occur much faster than the gas consumption due to star 
formation. Thus, we expect that the system would again break down in 
a hierarchy of well-separated equilibria, and the related equilibrium 
SFR would thus still be governed by the equilibrium
temperature in the cloud phase.

A new effect that arises for a clumpy ISM is another energy
dissipation process. In the case of a diffuse ISM, dissipation is mainly
driven by radiative cooling; however, in case of a system composed of
molecular clouds, inelastic cloud collisions provide a ``cooling''
mechanism dissipating the kinetic energy related to the motion of the
clouds.  Though the scaling of this dissipation process is identical
to radiative cooling (i.e.\ $\propto g^2$), the related timescales
differ strongly.  In general, cloud-cloud collisions operate on a much
longer timescale than radiative cooling or even the dynamics. Thus,
the dynamical evolution of a cloud system might operate not only on a
dynamical timescale, but also on a dissipational timescale which then
directly affects the SFR. We will discuss the
implications in a forthcoming paper in detail.

\noindent
{\bf Heating.}
For the heating we only considered massive stars and instantaneous
feedback. Though massive stars are considered to be a major source of
ISM heating, e.g.\ with respect to turbulence, other sources might be
interesting, too. For example, type I supernovae provide a nonlocal and a
non-instantaneous energy source. Especially, the delayed energy
feedback might be an important energy source after star formation
has ceased.  Similarly, the energy injected by an AGN might be an
important energy source, too.

%##########################################
%            Conclusions
%##########################################
\section{Summary and conclusions}
\label{sect_summary}

We investigated the evolution of star-forming dwarf galaxies by means
of an extended one-zone model. In contrast to previous one-zone models,
the dynamics of the galaxy has been taken into account by an
additional equation of motion for the mean radius of the galaxy.
Comparison with SPH models shows that this rather approximate
treatment is reasonable and provides a sensible description.  This
approach allows for a coupling between the dynamical state of the
galaxy and its internal properties, such as star formation activity and
the thermal state of the interstellar gas. Here we focussed especially
on the conditions under which starburst episodes could occur in
isolated galaxies.

We find that the seemingly complicated system of equations for the
numerous physical variables (radius, radial speed, gas mass, star
mass, thermal energy) breaks down into quite a clearly separated
system for the internal conditions of the galaxy and the dynamical
aspects that evolve more slowly. Thus, star formation and other
internal processes are strongly governed by the dynamical state. In
particular, the total SFR follows the
evolution of the volume directly, if it depends nonlinearly on the local gas
density. On the other hand, the dynamics are only very weakly
influenced by the internal state, because the gas temperature remains
well below virial temperature, primarily because of the strong rise of the
radiative cooling function near $10^4$ K.  The consequence is that the
evolution of the dynamical state of the galaxy remains fairly robust
against changes and details in the internal condition, specifically
the details of the prescriptions for star formation. 
We also expect that the inclusion of a multiphase description
of the ISM would not upset this characteristic.

The evolution of the total SFR is thus dominated by
the damped virial oscillations that cause enhancements of the SFR
whenever the galaxy is in a more compressed state. This main type of
quasi-periodic starbursts is related to the dynamical timescale of the
system. Typically, the star formation variations follow the variations
in the gas density induced by decaying virial oscillations. Because of the
short heating and cooling timescales the established SFR
remains close to the equilibrium SFR determined
by the current mean gas density.  Modifications in the adopted star
formation modes, i.e.\ assuming both spontaneous and induced star
formation/positive feedback, give qualitatively similar results.
Induced star formation neither leads to a fundamentally different
response nor causes a global conflagration. Variations in the IMF or
introduction of a time-variable IMF similarly do not change that
picture. In all cases, the variations in the SFR can be appreciable,
amounting to peak values up to 10 times the average value.

A second type of burst occurs for low gas densities and temperatures
beyond $10^4$\,K. This mode is not related to the dynamical evolution,
but an instability operating when the cooling function drops with
increasing temperature. The latter works for some temperature regimes
(related to H and He line emission) above $10^4$ K, which can only be
reached when there is a very low gas density.

Since the first type of bursts is governed by the evolution of the
dynamical state of the galaxy, it also seems plausible to discuss the
results of our simplified models for isolated galaxies in other
contexts, too. Virial oscillations following structural changes
induced by galaxy interactions would also lead to an enhancement of
the SFR, thus providing an additional mechanism that
shapes starbursts. Alternatively, if an interaction does not result in
a substantial mass redistribution, self-regulation might prevent a
starburst.

\begin{acknowledgements}
  We thank Jay Gallagher, Simone Recchi, and Shu-ichiro Inutsuka for
  stimulating discussions on starbursts and induced star formation.
  We are very grateful to the referee, Curt
    Struck, for detailed and fruitful comments that greatly improved
    the paper.  CT is grateful to the Observatoire de Strasbourg for
  hospitality and support within the visitor programme.
\end{acknowledgements}

%------------------------------------------------------------------
%                References
%------------------------------------------------------------------


\begin{thebibliography}{}

  \bibitem[2008]{bekki08}
   Bekki, K. 2008, \mnras, 338, L10

  \bibitem[2002]{berczik02}
   Berczik, P., Hensler, G., Theis, Ch., \& Spurzem, R. 2002, \apss, 281, 297

  \bibitem[1987]{binney87}
   Binney, J., \& Tremaine, S. 1987, Galactic Dynamics, Princeton Univ.\ Press

    \bibitem[2004]{blitz04}
   Blitz, L., Rosolowsky, E. 2004, \apj, 612, L29 

  \bibitem[1989]{boehringer89}
   B\"ohringer, H., Hensler, G. 1989, \aap, 215, 147

  \bibitem[1995]{burkert95}
   Burkert, A. 1995, \apj, 447, L25

  \bibitem[2006]{buyle06}
   Buyle, P., Michielsen, D., De Rijcke, S., et al., 2006, \apj, 649, 163 

  \bibitem[1972]{dalgarno72}
   Dalgarno, A., McCray, R.A. 1972, \araa, 10, 375

  \bibitem[2007]{dimatteo07}
   Di Matteo, P., Combes, F., Melchior, A.-L., Semelin, B. 2007, \aap, 468, 61

  \bibitem[2002]{dohmpalmer02}
   Dohm-Palmer, R.C., Skillman, E., Mateo, M., et al.\ 2002, \aj, 123, 813

  \bibitem[1997]{ehlerova97}
   Ehlerov\'a, S., Palou\v s, J., Theis, Ch., Hensler, G. 1997, \aap, 328, 121

  \bibitem[1984]{gallagher84}
   Gallagher, J.S., Hunter, D. 1984, \araa, 22, 37

  \bibitem[1978]{gerola78}
   Gerola, H., Seiden, P.E. 1978, \apj, 223, 129

  \bibitem[1993]{greggio93}
   Greggio, L., Marconi, G., Tosi, M., Focardi, P. 1993, \aj, 105, 894

  \bibitem[2006]{harfst06}
   Harfst, S., Theis, Ch., Hensler, G. 2006, \aap, 449, 509

  \bibitem[2004]{hensler04}
   Hensler, G., Theis, Ch., Gallagher, J. 2004, \aap, 426, 25

  \bibitem[2005]{hosokawa05}
   Hosokawa, T., Inutsuka, S. 2005, \apj, 623, 917

  \bibitem[2006]{hosokawa06}
   Hosokawa, T., Inutsuka, S. 2006, \apj, 646, 240

%  \bibitem[2006]{hosokawa06b}
%   Hosokawa, T., Inutsuka, S. 2006, \apj, 648, L131

  \bibitem[1983]{ikeuchi83}
   Ikeuchi, S., Tomita, H. 1983, \pasj, 35, 77

%\bibitem[2004]{kobayashi04}
%Kobayashi, M.A.R., Kamaya, H. 2004, \aap, 425, L41

  \bibitem[1995]{koeppen95}
   K\"oppen, J., Theis, Ch., Hensler, G. 1995, \aap, 296, 99, KTH95

  \bibitem[1998]{koeppen98}
   K\"oppen, J., Theis, Ch., Hensler, G. 1998, \aap, 331, 524

  \bibitem[2006]{koeppen07}
%   K\"oppen, J., Weidner, C., Kroupa, P. 2006, in preparation
   K\"oppen, J., Weidner, C., Kroupa, P. 2007, \mnras, 375, 673

  \bibitem[2004]{nikolic04}
   Nikolic, B., Cullen, H., Alexander, P. 2004, \mnras, 355, 874

  \bibitem[2004]{pelupessy04}
   Pelupessy F.I., van der Werf, P.P., Icke, V. 2004, \aap, 422, 55

  \bibitem[1992]{press92}
   Press W.H., Teukolsky S.A., Vetterling W.T., Flannery B.P.,
   1992, Numerical Recipes, Cambridge University Press, Cambridge

  \bibitem[2008]{quillen08}
   Quillen, A.C., Bland-Hawthorn, J. 2008, \mnras, 386, 2227

  \bibitem[1997]{samland97}
   Samland, M., Hensler, G., Theis, Ch.\ 1997, \apj, 476, 544

  \bibitem[1996]{sanders96}
   Sanders, D.B., Mirabel, I.F. 1996, \araa, 34, 749

  \bibitem[1986]{scalo86}
   Scalo, J., Struck-Marcell, C. 1986, \apj, 301, 77

  \bibitem[2005]{struck05} 
   Struck, C. 2005, in Proc.\ of {\it Starbursts: From 30 Doradus to
     Lyman Break Galaxies}, Cambridge, UK, R. de Grijs \& R.M. 
     Gonzalez-Delgado (eds.), \apss, 329, 163

  \bibitem[2006]{struck06} 
   Struck, C. 2006, in Astrophysics Update 2, J. Mason (ed.), p.\ 115

  \bibitem[1992]{theis92}
   Theis, Ch., Burkert, A., Hensler, G. 1992, \aap, 265, 465

  \bibitem[1993]{theis93}
   Theis, Ch., Hensler, G. 1993, \aap, 280, 85

  \bibitem[2005]{weidner05}
   Weidner, C., Kroupa, P. 2005, \apj, 625, 754
 
  \bibitem[2001]{vanzee01}
   van Zee, L. 2001, \aj, 121, 2003

\end{thebibliography}
\end{document}